\documentclass[10pt,aps,pre,showpacs,twocolumn,groupedaddress,longbibliography]{revtex4-2}
\usepackage{graphicx}
\usepackage{latexsym}
\usepackage{amssymb}
\usepackage{bbold}
\usepackage{color}
\usepackage{bm}
\usepackage{amsmath}

\usepackage[pdftex,bookmarks,colorlinks,breaklinks]{hyperref} 
\hypersetup{linkcolor=red,citecolor=blue,filecolor=dullmagenta,urlcolor=blue} 
\hypersetup{pdftoolbar=true,pdfpagemode=UseNone,pdfstartview=FitH, colorlinks=true,extension=}

\newcommand{\onehalf}{$\frac{1}{2}$\;}
\newcommand{\unity}{\mathbb{1}}
\newcommand{\bi}[1]{\boldsymbol{#1}}
\newcommand{\mat}[4]{\left(\begin{array}{cc} #1 & #2\\ #3 & #4\\ \end{array}\right)}
\newcommand{\matfix}[4]{\left(\begin{array}{p{6ex}p{6ex}} \hfill $#1$ \hspace*{\fill} & \hfill $#2$\hspace*{\fill} \\ \hfill $#3$\hspace*{\fill} & \hfill $#4$\hspace*{\fill} \end{array}\right)}
\newcommand{\sigdt}{\mat{\sigma_z}{\cdot}{\cdot}{\sigma_z}}
\newcommand{\unit}{\unity_{_4}}
\newcommand{\sigx}{\mat{\cdot}{1}{1}{\cdot}}
\newcommand{\sigy}{\mat{\cdot}{-i}{i}{\cdot}}
\newcommand{\sigz}{\mat{1}{\cdot}{\cdot}{-1}}
\newcommand{\sigP}{\mat{\cdot}{i\sigma_y}{-i\sigma_y}{\cdot}}
\newcommand{\sigD}{\mat{\cdot}{\unity}{\unity}{\cdot}}
\newcommand{\sigwP}{\mat{\sigma_z}{\cdot}{\cdot}{-\sigma_z}}
\newcommand{\sigwD}{\mat{\sigma_z}{\cdot}{\cdot}{\sigma_z}}
\newcommand{\sigmaP}{\bm{\sigma_P}}
\newcommand{\tsigmaP}{\bm{\tilde{\sigma}_P}}
\newcommand{\sigmaD}{\bm{\sigma_D}}
\newcommand{\tsigmaD}{\bm{\tilde{\sigma}_D}}
\newcommand{\sigmaa}{\bm{\sigma_a}}
\newcommand{\tsigmaa}{\bm{\tilde{\sigma}_a}}
\newcommand{\sigmab}{\bm{\sigma_b}}
\newcommand{\tsigmab}{\bm{\tilde{\sigma}_b}}

\newcommand{\bh}{\bi{h}}
\newcommand{\bF}{\bi{F}}
\newcommand{\biG}{\bi{G}}
\newcommand{\bG}{G}
\newcommand{\bK}{\bi{K}}
\newcommand{\bS}{S}

\newcommand{\bV}{\bi{V}\!\!}
\newcommand{\ba}{a}
\newcommand{\bb}{b}
\newcommand{\bphi}{\bm{\varphi}}
\newcommand{\bPsi}{\bm{\Psi}}

\graphicspath{{./figs/},{./}}

\begin{document}
\title{Realization of an NMR analog in a microwave network with symplectic symmetry}

\author{Finn Schmidt}
\affiliation{Fachbereich Physik der Philipps-Universit\"{a}t Marburg, D-35032 Marburg, Germany}
\author{Tobias Hofmann}
\affiliation{Fachbereich Physik der Philipps-Universit\"{a}t Marburg, D-35032 Marburg, Germany}
\author{Ulrich Kuhl}
\email{ulrich.kuhl@univ-cotedazur.fr}
\affiliation{Universit\'{e} C\^{o}te d'Azur, CNRS, Institut de Physique de Nice (INPHYNI), 06108 Nice, France, EU}
\affiliation{Fachbereich Physik der Philipps-Universit\"{a}t Marburg, D-35032 Marburg, Germany}
\author{H.-J.\ St\"{o}ckmann}
\email{stoeckmann@physik.uni-marburg.de}
\affiliation{Fachbereich Physik der Philipps-Universit\"{a}t Marburg, D-35032 Marburg, Germany}
\date{\today}

\begin{abstract}
In a previous paper, we realized a microwave network with symplectic symmetry simulating a spin 1/2 (Rehemanjiang et\,al.\ [Phys.\ Rev.\ Lett.\ 117, 064101 (2016)]), following a suggestion by Joyner et\,al.\ [Europhys.\ Lett.\ 107, 50004(2014))].
The network consisted of two identical sub-units coupled by a pair of bonds with a length difference corresponding to a phase difference of $\pi$ for the waves traveling through the bonds.
In such a symmetry each eigenvalue appears as a two-fold degenerate Kramers doublet.
Distorting the symmetry the degeneracy is lifted which may be interpreted in terms of the Zeeman splitting of a spin 1/2 in an external magnetic field.
In the present work, a microwave analog of a spin resonance is realized.
To this end, two magnetic fields have to be emulated, a static and a radio-frequency one.
The static one is realized by detuning the length difference from the $\pi$ condition by means of phase shifters, the radio-frequency field by modulating the length difference of another pair of bonds by means of diodes with frequencies up to 125\,MHz.
Features well-known from magnetic resonance such as the transition from the laboratory to the rotating frame, and Lorentzian shaped resonance curves can thus be realized.
\end{abstract}

\maketitle

\section{Introduction}
\label{sec:intro}

\subsection{Background}

Random matrix theory has proven very successful in the description of the universal features of the spectra of chaotic systems \cite{meh91,haa18}.
Depending on the presence or absence of time-reversal symmetry (TRS) there are three ensembles, the orthogonal one with TRS and no spin \onehalf, the unitary one for systems without TRS, and the symplectic one for systems with TRS and a spin \onehalf.
There are numerous experimental studies of the spectral statistics of first two ensembles, see Ref.~\cite{stoe22} for a recent review.
But for the realization of the symplectic ensemble one had to wait for many years because of the difficulty to realize a spin \onehalf.
The breakthrough finally came from Joyner and coworkers \cite{joy14} who proposed a graph with a symplectic symmetry without a spin \onehalf.
This idea was experimentally realized by us \cite{reh16,reh18} and thereafter by others \cite{lu20,ma23,law24} in microwave networks.

The proposed network, called graph in the following, consisted of two subgraphs, one the complex conjugate of the other, realized in the experiment by circulators with opposite sense of rotation.
A circulator is a microwave device introducing directionality:
Microwaves entering through ports 1, 2, 3 exit via ports 2, 3, 1, respectively.
The second ingredient is a pair of bonds coupling, e.\,g., vertices $n$ and $m$ in one subgraph with the symmetry equivalent ones $\bar{m}$ and $\bar{n}$, in the other one.
For a symplectic symmetry to hold there must be a phase difference of $\pi$ (or any odd integer multiple of $\pi$) for the waves propagating through the two bonds.
In the spirit of the spin analogy one subgraph corresponds to the spin-up and the other subgraph to the spin-down component, and the phase difference of $\pi$ between the two bonds mimics the fact that a spinor rotation by $2\pi$ changes the sign of the spinor.

\begin{figure}
  \includegraphics[width=\linewidth]{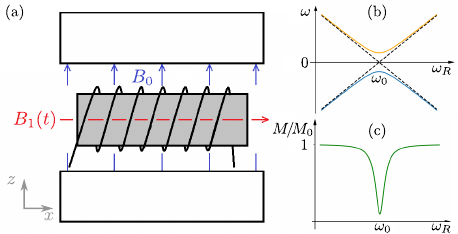}\\
  \caption{ \label{fig:NMR}
    (a) Sketch of a typical NMR spectrometer.
    (b) Eigenvalues of a spin \onehalf, exposed to a static magnetic field in $z$ and a radio-frequency field in $x$ direction, in the rotating coordinate system, and
    (c) magnetization of the sample in dependence of the angular frequency $\omega_R$ of the radio-frequency field.
  }
\end{figure}

In this paper we take this analogy literally and ask the question:
If there is a spin analog in a network with symplectic symmetry, is there, perhaps, also an analog to spin resonance?

\subsection{Basics of nuclear magnetic resonance}

In a typical nuclear magnetic resonance (NMR) experiment \cite{sli80} a probe, containing an ensemble of nuclear spins $\vec{I}$, is exposed to a static magnetic field $B_0$, by convention in $z$ direction, and a radio-frequency field $B_1(t)=2B_1\cos\omega_Rt$ in $x$ direction, generated by a coil wrapping the probe, see Fig.~\ref{fig:NMR}(a).
The system is described by the time-dependent Schr\"odinger equation
\begin{equation} \label{eq:schr}
  \dot{\psi} = -\frac{i}{\hbar} H_\mathrm{NMR} \psi
\end{equation}
with the Hamiltonian
\begin{equation} \label{eq:HNMR}
  H_\mathrm{NMR} = -\hbar \left(\omega_0I_z + 2\omega_1\cos\omega_Rt I_x\right)\,,
\end{equation}
where $\omega=\gamma B$ is the Larmor angular frequency, with the gyromagnetic ratio $\gamma$.
The factor $2$ has been introduced for later convenience.
For a spin $I$=\onehalf, assumed in the following, the angular momentum operators may be expressed in terms of the Pauli matrices, $\vec{I}=\frac{1}{2}\vec{\sigma}$, with
\begin{equation} \label{eq:Pauli}
  \sigma_x=\sigx\,,\,
  \sigma_y=\sigy\,,\,
  \sigma_z=\sigz\,.
\end{equation}
The time-dependent part in Eq.~(\ref{eq:HNMR}) may be interpreted in terms of the interaction of a spin with a superposition of two magnetic fields rotating clock and counter-clockwise in the $xy$ plane.
The standard approach to solve the Schr\"odinger equation is a transformation into a rotating coordinate system to remove the time-dependency.
With the rotated wave function,
\begin{equation} \label{eq:psiR}
  \psi_R = e^{-i\frac{\omega_R}{2}t\sigma_z} \psi
\end{equation}
the Schr\"odinger equation is transformed into
\begin{equation} \label{eq:scrR}
  \dot{\psi_R} = \frac{i}{2} \left[(\omega_0-\omega_R)\sigma_z+2\omega_1\cos\omega_Rt (\sigma_x)_R\right]\psi_R
\end{equation}
with
\begin{eqnarray} \label{eq:sigmaR}
  2\cos\omega_Rt (\sigma_x)_R
  &=& 2\cos\omega_Rt\, e^{i\frac{\omega_R}{2}t\sigma_z}\sigma_x e^{-i\frac{\omega_R}{2}t\sigma_z} \nonumber \\
  &=& \sigma_x + \cos 2\omega_Rt \sigma_x -\sin 2\omega_Rt \sigma_y\,.
\end{eqnarray}
Due to the transformation into the rotating system the field component rotating synchronously with the coordinate system has become static, whereas the other component now rotates with the double frequency in the opposite direction.
Ignoring this component, a standard practice in NMR, the Hamiltonian has become static in the rotating system,
\begin{equation} \label{eq:HNMRR}
  H_\mathrm{NMR}^R = -\frac{1}{2} \left[(\omega_0-\omega_R)\sigma_z + \omega_1\sigma_x \right]
\end{equation}
The eigenvalues of $H_\mathrm{NMR}^R$ are given by
\begin{equation} \label{eq:eigenR}
  \omega_\pm = \pm\frac{1}{2} \sqrt{(\omega_0-\omega_R)^2 + \omega_1^2}\,,
\end{equation}
see Fig.~\ref{fig:NMR}(b).
In a standard NMR experiment the magnetization $M$, proportional to the spin polarization, is studied as a function of $\omega_0$ or $\omega_R$.
The avoided crossing exhibited by the eigenvalues in the rotating frame at $\omega_0=\omega_R$ implies a Lorentzian resonance curve,
\begin{equation} \label{eq:res}
  M/M_0 = 1 - \frac{\omega_1^2}{(\omega_0-\omega_R)^2 + \omega_1^2}\,,
\end{equation}
see Fig.~\ref{fig:NMR}(c), where $M_0$ is the equilibrium polarization, usually resulting from a Boltzmann polarization.

We shall follow exactly the same strategy in our approach to realize an NMR analog in a microwave network.
To this end we have to find analogs for two magnetic fields, a static one, which is easy, and a time-dependent one, which is a challenge.

A joint short paper is published as \cite{hof24}, where the focus is on the phenomenology and its experimental realization.
In this paper we present a rigorous mathematical treatment proving to be surprisingly complex, as well as further experimental results.

\begin{figure}
  \includegraphics[width=\linewidth]{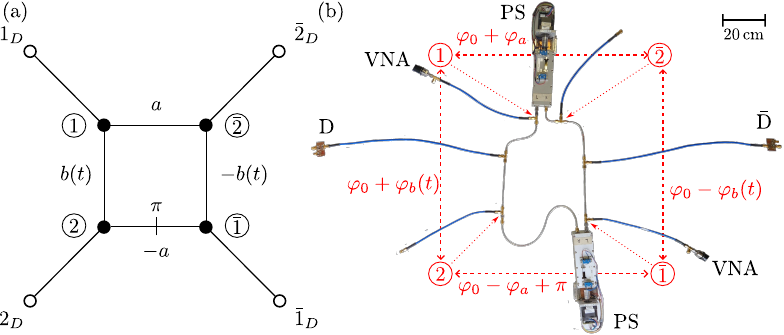} 
  \caption{ \label{fig:sketch}
    (a) Sketch of the idealized graph,
    (b) Photograph of the experimental set-up.
  }
\end{figure}

\section{The experiment}
\label{sec:experiment}

\subsection{The set-up} \label{sec:expgraph}

Any perturbation of symplectic symmetry implies a lifting of the Kramers degeneracy which may be interpreted in terms of a Zeeman splitting in an effective magnetic field as will be discussed in detail below.
In our previous work \cite{reh16,reh18} this was achieved by changing the length difference of $\Delta l=l_1- l_2$ of the two bonds connecting the two subgraphs, thereby detuning the phase difference $\Delta\varphi=k\Delta l$ for the waves propagating through the pair of bonds from an odd integer multiple of $\pi$, needed for the symplectic symmetry.
The realization of an NMR equivalent needs the emulation of two magnetic fields orthogonal to each other.
Our previous studies suggested to do this by means of two pairs of bonds, where the deviations of the two phase differences $\Delta\varphi_0\sim\omega_0$ and $\Delta\varphi_1\sim\omega_1$ from the $\pi$ condition may be interpreted in terms of magnetic fields.
The diabolo presented in \cite{hof24}, found for the energy surface of a Kramers doublet in dependence of $\Delta\varphi_0$ and $\Delta\varphi_1$, was obtained with such a geometry.
There are, however, serious disadvantages with this approach.
First, one has to fulfill the $\pi$ condition for two pairs of bonds at the same time meaning a very careful adjustment of the cable lengths, and second, the ``magnetic fields'' realized in such a way are not automatically orthogonal.
Therefore in the present work we used a system, lacking these disadvantages, with a symmetry which had already been used by us for a realization of the chiral symplectic ensemble \cite{reh20}, not in graphs, however, but by coupled dielectric resonators.

Figure~\ref{fig:sketch}(a) shows a sketch of the idealized graph.
The unperturbed graph consists of four nodes $1,2,\bar{1}, \bar{2}$, obeying Neumann boundary conditions, coupled by four bonds of equal length $l$ to form a square.
At each of the nodes a dangling bond is attached, again of length $l$, and terminated by a short end corresponding to a Dirichlet boundary condition.
Along the bond $\bar{1}2$ the wanted phase shift of $\pi$ is applied.
The symplectic symmetry is detuned by applying static relative length changes $\pm a$ of opposite sign at bonds $1\bar{2}$ and $\bar{1}2$.
Along bonds $12$ and $\bar{1}\bar{2}$ time-dependent relative length changes $\pm b(t)$ are applied.
The system shown in Fig.~\ref{fig:sketch}(a) means an idealization, which cannot be realized experimentally one-to-one, but it allows for an exact analytic treatment, essential for an understanding of what is going on.

Figure~\ref{fig:sketch}(b) shows a photograph of the actual experimental set-up.
The graph has the same geometry as the idealized one, but there are some differences in detail.
The geometrical length of the individual bonds amounted to $l_\mathrm{geo}= 30.38$\,cm, corresponding to an optical length of $l=l_\mathrm{opt} = nl_\mathrm{geo} = 43.75$\,cm, where $n=1.44$ is the index of refraction.
All lengths given in the following refer to the optical ones.
The phase shift of $\pi$ was realized by adding an extra length to the respective bond corresponding to half of the wavelength of the selected resonance.
The fine tuning was performed by means of microwave devices called phase shifters.
In reality they do not change the phase but the length (just like a trombone).
The same phase shifters have been used for the detuning of the phase difference for the $1\bar{2}/\bar{1}2$ pair from the $\pi$ condition.
Note that there are no circulators breaking time-reversal symmetry in the two subgraphs.
Circulators always introduce considerable absorption which showed up to be intolerable for the present studies.
For a symplectic symmetry a break of time reversal symmetry is not really needed, it is sufficient that the two subgraphs are complex conjugates of each other, which is automatically the case if both of them are real.

\subsection{The static perturbation}

\begin{figure}
  \includegraphics[width=\linewidth]{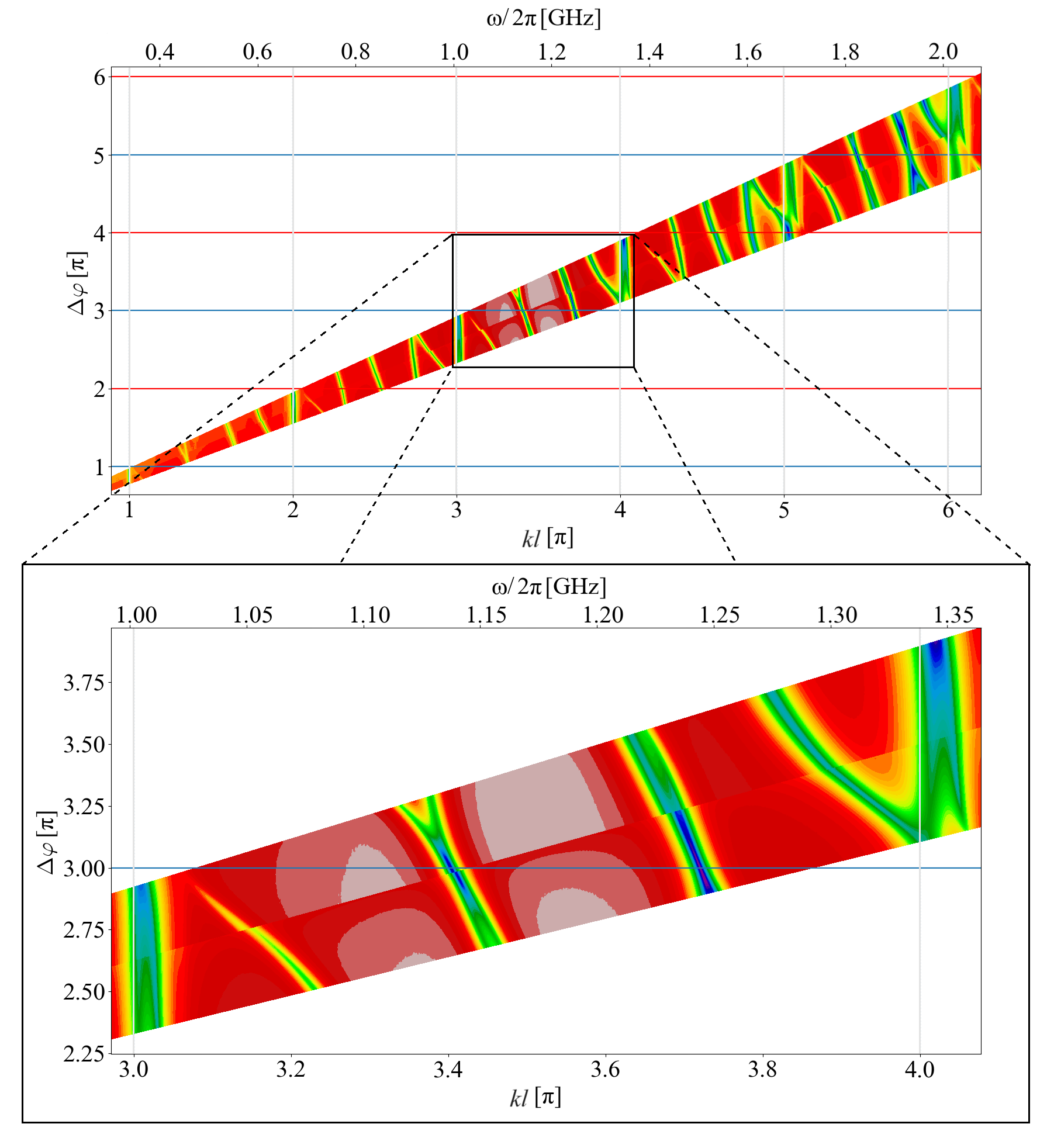}
  \caption{ \label{fig:pilines}
    Reflection spectra as a function of frequency $\nu=\omega/2\pi$ for the graph shown in Fig.~\ref{fig:sketch}(b) for different phase differences $\Delta\varphi=k\Delta l$ ($\Delta l$:
    difference of the lengths of bonds $1\bar{2}$ and $\bar{1}2$).
    Each resonance appears in a light color on a dark background.
    The abscissa label above the plot shows the frequency $\nu=\omega/2\pi$, the label below the plot the corresponding $k$ value.
    The Kramers doublet at 1.12\,GHz had been used for the time-dependent studies.
 }
\end{figure}

In a preliminary study the spectrum of the graph was determined in dependence of the phase difference $\Delta\varphi$ for the $1\bar{2}/\bar{1}2$ pair, and no detuning for the $12/\bar{1}\bar{2}$ pair.
The length of bonds $1\bar{2}$ was changed by the phase shifters step-wise to vary $\Delta\varphi$.
Using the relation $\Delta\varphi=k\Delta l$ the length shift $\Delta l$ was translated into a corresponding phase shift.
A detailed description of the technique can be found in section 2.1.3.\ of Ref.~\cite{stoe22}.

Figure~\ref{fig:pilines} shows spectra of the reflection $S_{11}$ in dependence of $\Delta\varphi$ in a color plot.
The resonances appear in green on a red background.
At the odd integer multiples of $\pi$, depicted by horizontal blue lines, the graph is symplectically symmetric, and all resonances appear as degenerate Kramers doublets.
Any deviation from the $\pi$ lines means a splitting of the Kramers doublets which may be interpreted in terms of a Zeeman splitting in a magnetic field.
A detailed interpretation of the resonances will be given in section \ref{sec:secular}.

The wedge shape of the experimentally explored region reflects the limitation of the phase shifter.
It allows for a limited length change by $\Delta l=4$\,cm only, corresponding to a $\Delta\varphi$ range of $0.1\dots0.3 \,\pi$ for the shown frequency range.
To increase the accessible range, additional pieces of fixed lengths had been introduced.
The figure shows two successive wedges such obtained.
The borderline between the two measurements is clearly seen in the enlarged region in the lower plot.

The limitations resulting from the range of the phase shifter are less stringent as it may seen, since the patterns are periodic with respect to both axes.
First, $\Delta\varphi$ obviously is $2\pi$ periodic.
And since all bonds have the same length $l$, $k$ is periodic with a period of $\pi/l$.
The limits of the $k$ windows are indicated in the plot by vertical grey lines.

\subsection{The time-dependent perturbation}

The experimental realization of time dependence was the main challenge of the present work.
The total optical length of the graph was $l_\mathrm{tot}=4.33$\,m, including the extra length to realize the $\pi$ phase shift.
This corresponds to a mean spacing of $70$\,MHz of the Kramers doublets.
Typical line widths of the resonances due to absorption and coupling amounted to about $10$\,MHz.
Hence both Zeeman splittings and the frequencies of the time variations must be in this range.
For the time variations we used diodes allowing for bipolar switches between {\em open} and {\em short} up to $125$\,MHz.
We are not aware of existing diodes allowing for periodic variations in this frequency range.
Thus there are always unwanted excitations also at the higher frequency harmonics.
We did not see, however, any influence of the higher harmonics in the experiments.
We tried a number of alternatives to implement the diodes, and finally came to the solution to attach dangling bonds in the center of the $12$ and $\bar{1}\bar{2}$ bonds via T junctions with diodes at the end allowing for a periodic switch of the lengths of the dangling bonds, with a length change $\Delta l$ of the order of some centimeter.
This does not change explicitly the lengths of the two bonds, but it changes the phase a wave acquires when propagating through the bonds, which for a given frequency means just the same as a change of the length.
The switches were performed anti-cyclically with a positive $\Delta l$ for one bond accompanied by a negative $\Delta l$ for the other one, thus preserving the mean density of states.
To this end two types of diodes had been used differing in their polarities.
Details can be found in appendix~\ref{app:diode}.

\section{Theory}
\label{sec:theory}

\subsection{The continuity equation}

There are two constraints governing wave functions in graphs.
First, energy conversation implies that all wave functions meeting at a node $n$ have the same value $\varphi_n(t)$ at this node.
The usual separation of the time dependence is not possible here, since the graph is time dependent.
The wave function connecting nodes $n$ and $m$ with a bond of length $l_{nm}$ may be written as
\begin{equation} \label{eq:psi}
  \psi_{nm}(x,t) = \varphi_{nm}(x,t) + \varphi_{mn}(l_{nm}-x,t)\,,
\end{equation}
where the $\varphi_{nm}(x,t)$ satisfy the boundary conditions
\begin{equation} \label{eq:phi}
  \varphi_{nm}(0,t) = \varphi_n(t)\,, \quad \varphi_{nm}(l_{nm},t) = 0\,,
\end{equation}
whence follow the boundary conditions $\psi_{nm}(0,t)=\varphi_n(t)$ and $\psi_{nm}(l_{nm},t)=\varphi_m(t)$ for $\psi_{nm}(x,t)$ as demanded by energy conservation.

The $\varphi_{nm}(x,t)$ obey the time-dependent wave equation.
In the calculations variations of length $l$ would be inconvenient.
But since the $\varphi_{nm}(x,t)$ depend on the {\em optical} length $l=nl_\mathrm{geo}$ only, a change of $l_\mathrm{geo}$ can be substituted by a corresponding change of $n$.
The $\varphi_{nm}(x,t)$ then are solutions of a correspondingly rescaled wave equation,
\begin{equation} \label{eq:wave}
  \left[\square + 2v_{nm} \frac{\partial^2}{\partial x^2}\right] \varphi_{nm}(x,t) = 0\,,
\end{equation}
where
\begin{equation} \label{eq:square}
  \square = -\frac{1}{c^2} \frac{\partial^2}{\partial t^2} + \frac{\partial^2}{\partial x^2}
\end{equation}
is the d'Alembert operator in one dimension, $c=c_0/n$ is the velocity of light in the bonds, and the $v_{nm}$ are given by $v_{nm}=0$ for the dangling bonds, $v_{1\bar{2}/\bar{1}2}= \pm a$, $v_{12/\bar{1}\bar{2}}= \pm b(t)$.
Here $a$ and $b(t)=2b\cos(\omega_Rt)$ are the relative length changes due to phase shifter and diode, respectively.
Throughout this paper it is assumed that these changes are small and can be treated in first order.

We arrange the $\varphi_{nm}$ in terms of spinor-like column vectors,
\begin{equation} \label{eq:Phi}
  \bphi_B = \left(\begin{array}{r}
    \varphi_{11_D} \\ \varphi_{\bar{1}\bar{1}_D} \\ \varphi_{2\bar{2}_D} \\ \varphi_{\bar{2}2_D} \\
  \end{array}\right)\,,\quad\!
  \bphi_P = \left(\begin{array}{r}
    \varphi_{1\bar{2}} \\ \varphi_{\bar{1}2} \\ \varphi_{2\bar{1}} \\ \varphi_{\bar{2}1} \\
  \end{array} \right)\,,\quad\!
 \bphi_D = \left(\begin{array}{r}
    \varphi_{12} \\ \varphi_{\bar{1}\bar{2}} \\ \varphi_{21} \\ \varphi_{\bar{2}\bar{1}} \\
    \end{array} \right)\,,
\end{equation}
where the subscripts refer to dangling bonds (B), phase shifters (P), and diodes (D), respectively.
The wave functions connecting the nodes may compactly be written in terms of the $\varphi_A(x,t)$ as
\begin{eqnarray} \label{eq:Psi}
  \bPsi_B(x,t) &=& \bphi_B(x,t)\,, \nonumber\\[1ex]
  \bPsi_P(x,t) &=& \left( \begin{array}{r}
    \varphi_{1\bar{2}} \\ \varphi_{\bar{1}2} \\ \varphi_{2\bar{1}} \\ \varphi_{\bar{2}1} \\
    \end{array} \right)(x,t)+
  \left( \begin{array}{r}
    \varphi_{\bar{2}1} \\ -\varphi_{2\bar{1}} \\ -\varphi_{\bar{1}2} \\ \varphi_{1\bar{2}} \\
  \end{array} \right)(l-x,t)\nonumber\\
  &=& \bphi_P(x,t)+\sigP\bphi_P(l-x,t)\,,\nonumber\\[1ex]
  \bPsi_D(x,t) &=& \left( \begin{array}{c}
    \varphi_{12} \\ \varphi_{\bar{1}\bar{2}} \\ \varphi_{21} \\ \varphi_{\bar{2}\bar{1}} \\
  \end{array} \right)(x,t)+
  \left( \begin{array}{r}
    \varphi_{21} \\ \varphi_{\bar{2}\bar{1}} \\ \varphi_{12} \\ \varphi_{\bar{1}\bar{2}} \\
  \end{array} \right)(l-x,t)\nonumber\\
  &=& \bphi_D(x,t)+\sigD\bphi_D(l-x,t)\,,
\end{eqnarray}
where Eq.~(\ref{eq:psi}) has been used.
The minus signs in the second column vector contributing to $\bPsi_P(x,t)$ are a consequence of the phase shift of $\pi$ along the bond $\bar{1}2$.

The second constraint is current conservation \cite{kot99a}, resulting in a continuity equation for each node.
It may be compactly expressed in terms of the $\bPsi_A(x,t)$ as
\begin{equation} \label{eq:cont} \nonumber
  \bPsi_B'(0,t) + \bPsi_P'(0,t) + \bPsi_D'(0,t) = 0
\end{equation}
or
\begin{gather} \label{eq:cont1}
  \bphi_B'(0,t) + \bphi_P'(0,t) + \bphi_D'(0,t) \nonumber\\ - \sigmaP\bphi_P'(l,t) - \sigmaD\bphi_D'(l,t)
  = 0\,,
\end{gather}
where
\begin{equation} \label{eq=sigmaPD}
  \sigmaP = \sigP\,, \quad \sigmaD = \sigD\,.
\end{equation}
Equation~(\ref{eq:cont}) holds for Neumann boundary conditions, as realized, e.\,g., in standard T junctions.

\subsection{The static case}
\label{sec:theo_stat}

\subsubsection{The secular matrix}
\label{sec:secular}

For the column vectors~(\ref{eq:Phi}) the wave equation~(\ref{eq:wave}) may be compactly written as
\begin{equation} \label{eq:wave1}
 \left\{ \square\cdot\unit+2\bV_A\frac{\partial^2}{\partial x^2}\right\}\bphi_A=0
\end{equation}
($A=B$, $P$, $D$), where $\unit$ is the four-dimensional unit matrix, and
\begin{equation} \label{eq:VA}
  \bV_B=0\cdot\unit\,, \bV_P=a\sigmaa\,, \bV_D=b(t)\sigmab
\end{equation}
with
\begin{equation} \label{eq:sigmaab}
  \sigmaa=\sigwP\,,\quad \sigmab=\sigwD\,.
\end{equation}
For a time-independent perturbation $b(t)=b=\mathrm{const.}$ the time dependence can be separated,
\begin{equation} \label{eq:tsep}
  \bphi_A(x,t) = e^{-i\omega t} \bphi_A(x)\,.
\end{equation}
In the absence of any perturbation, $a=b(t)=0$,
the wave equation~(\ref{eq:wave}) is immediately solved yielding the same wavefunction for each bond,
\begin{equation} \label{eq:sol}
  \bphi_0(x) = \frac{\sin k(l-x)}{\sin kl}\bphi_0\,,
  \quad 
  \bphi_0 = \left(\begin{array}{c}
    \varphi_1 \\ \varphi_{\bar{1}} \\ \varphi_2 \\ \varphi_{\bar{2}} \\
  \end{array}\right)\,.
\end{equation}
For this case the continuity equation~(\ref{eq:cont}) reduces to
\begin{equation} \label{eq:cont2}
  \left[-3f\unit + g \mat{\cdot}{\unity+i\sigma_y}{\unity-i\sigma_y}{\cdot}\right] \bphi_0 = 0\,,
\end{equation}
where
\begin{equation} \label{eq:fg}
  f = k\cot(kl)\,, \quad g = k/\sin(kl)\,,
\end{equation}
or
\begin{equation} \label{eq:sec}
  \bh_0\bphi_0 = 0
\end{equation}
with the secular matrix
\begin{equation} \label{eq:h0}
  \bh_0 = \left(\begin{array}{cc}
    -3f\, \unity & g\,(\unity+i\sigma_y) \\
    g\,(\unity-i\sigma_y) & -3f\, \unity \\
  \end{array}\right)\,.
\end{equation}
Next $\bh_0$ is diagonalized.
This is achieved by means of the transformation
\begin{equation} \label{eq:trans}
  \tilde{\bphi}_0 = \tilde{\bi{R}}\bphi_0\,,
  \quad 
  \tilde{\bh}_0 = \tilde{\bi{R}}\bh_0\tilde{\bi{R}}^\dag\,,
\end{equation}
where
\begin{equation} \label{eq:rot}
  \tilde{\bi{R}} = \frac{1}{\sqrt{2}}
  \left( \begin{array}{cc}
    \unity & i\sigma_y\\
    i\sigma_y & \unity\\
  \end{array}\right)
  \left(\begin{array}{cc}
    \varepsilon_y & \cdot \\
    \cdot & \varepsilon_y^\dag \\
  \end{array}\right)\,,
  \quad
  \varepsilon_y = e^{i\frac{\pi}{8}\sigma_y}\,.
\end{equation}
By means of this transformation $\bh_0$ is turned into
\begin{equation} \label{eq:h01}
  \tilde{\bh}_0=\left(
  \begin{array}{cc}
    (-3f+ g\sqrt{2})\, \unity & \cdot \\
    \cdot & (-3f- g\sqrt{2})\, \unity \\
  \end{array}
  \right)\,.
\end{equation}
In the absence of the perturbations there are, as expected, two two-fold degenerate zeros of $|\bh|$ at $3f=\pm\sqrt{2}g$, whence follows
\begin{equation} \label{eq:zeros}
  \tan(k_\pm l)=\pm\sqrt{\frac{7}{2}}\,,
\end{equation}
or $k_\pm l=\pm 0,344\pi+n\pi$, $n \in \mathbb{N}$.
Both continuity equation~(\ref{eq:cont}) and wave equation~(\ref{eq:wave1}) also hold in the new basis, if all occurring matrices are transformed correspondingly.
The matrices in the new basis are compiled in appendix~\ref{app:trans}.

Within each $k$ window we expect two Kramers doublets from Eq.~(\ref{eq:zeros}), meaning altogether four eigenvalues.
But there is a problem:
The mean density of state is $\bar{\rho}=l_\mathrm{tot}/\pi$, where $l_\mathrm{tot}=8l$ is the total length of the graph.
Thus there should be eight eigenvalues within each $k$ window.
Where are the missing four ones?
The answer comes from the spectral duality property of the secular matrix $h(k)$ \cite{hof21}:
The {\em zeros} of $h(k)$ constitute the spectrum of the graph, but there is another spectrum obtained from the {\em poles} of $h(k)$.
It corresponds to the superposition of the spectra of the individual bonds.
If all bond lengths are incommensurable, the two spectra are disjoint.
But for commensurable bond
lengths it may happen that some eigenvalues are at the same time members of both spectra.
In this case there is a cancellation of zeros in the nominator and denominator of $h(k)$, meaning that the eigenvalues in question are missing in the set of zeros of $h(k)$.
This type of eigenvalues has been termed ``topological'' by Gnutzmann et al.~\cite{gnu13}.
In the present case there are four of them:
one resonance living along the bonds $1_D1-12-22_D$ and being zero on all other bonds, as well as its three symmetry equivalents.
In the experiment the topological resonances appear at the boundaries of the $k$ windows, with a fourfold degeneracy at the odd integer $\pi$ lines, see Fig.~\ref{fig:pilines}.

The detuning of the bond lengths results in two extra terms in the secular matrix
\begin{equation} \label{eq:sec2}
  \bh=\bh_0+\bh_a+\bh_b\,.
\end{equation}
In the basis, where $h_0$ is diagonal, they are given by
\begin{eqnarray} \label{eq:hab}
  \tilde{\bh}_a &=& ak\left[
    \frac{f'}{\sqrt{2}} \left(\begin{array}{cc}
      \sigma_z & -\sigma_z \\
      -\sigma_z & -\sigma_z
    \end{array}\right)
    -g'\left(\begin{array}{cc}
      \sigma_z & \cdot \\
      \cdot & \sigma_z
    \end{array}\right)
  \right]\,, \nonumber\\
  \tilde{\bh}_b &=& bk\left[
    \frac{f'}{\sqrt{2}} \left(\begin{array}{cc}
      -\sigma_x & -\sigma_x \\
      -\sigma_x & \sigma_x
    \end{array}\right)
  +g' \left(\begin{array}{cc}
    \sigma_x & \cdot \\
    \cdot & \sigma_x
  \end{array}\right)
  \right]\,,
\end{eqnarray}
where $f'$, $g'$ denote the derivatives of $f$, $g$ with respect to $k$.
The derivation is presented in appendix~\ref{app:sec}.

For small perturbations the off-diagonal blocks of $\tilde{h}_a$ and $\tilde{h}_b$ may be discarded, and $\tilde{h}$ adopts block diagonal form with diagonal blocks $\tilde{h}_+$ and $\tilde{h}_-$, given by
\begin{equation} \label{eq:hp}
  \tilde{h}_\pm = (-3f \pm g\sqrt{2})\, \unity + k\left(\pm\frac{ f'}{\sqrt{2}} -g'\right)(a\sigma_z-b\sigma_x)\,.
\end{equation}

One remark may be appropriate:
The diagonalization (\ref{eq:rot}) of $\bh_0$ is not unique.
Every additional transformation $R_1=\text{\tiny$\mat{R_{1+}}{\cdot}{\cdot}{R_{1-}}$}$, not destroying the the block diagonal structure of $\tilde{\bh}_0$, is allowed.
This freedom has been used to obtain expressions for $\tilde{h}_a$ and $\tilde{h}_b$, which are particular suited for the treatment of the time-dependent case, with $\tilde{h}_a$ depending only on $\sigma_z$, and $\tilde{h}_b$ only on $\sigma_x$.

\subsubsection{The scattering matrix}

For the measurement the graph is connected via vertices $1$ and $\bar{1}$ to a vector network analyzer measuring reflection amplitudes $S_{11}$, $S_{\bar{1}\bar{1}}$ and transmission amplitudes $S_{1\bar{1}}$, $S_{\bar{1}1}$ between the ports.
The scattering matrix
\begin{equation} \label{eq:S}
  \bS = \left(\begin{array}{cc}
    S_{11} & S_{1\bar{1}} \\
    S_{\bar{1}1} &S_{\bar{1}\bar{1}}
 \end{array}\right)
\end{equation}
relates vectors $\ba$, $\bb$ of incoming and outgoing amplitudes via
\begin{equation} \label{eq:scatt}
  \bb = \bS\ba\,,
\end{equation}
where
\begin{equation} \label{eq:ab}
  \ba(\omega,t) = e^{-i\omega t}\left(\begin{array}{c}
    a_1(\omega) \\
    a_{\bar{1}}(\omega) \\
  \end{array}\right)\,,\quad
  \bb(\omega,t) = e^{-i\omega t}\left(\begin{array}{c}
    b_1(\omega) \\
    b_{\bar{1}}(\omega) \\
  \end{array}\right)\,.
\end{equation}
For the static case we can apply time-independent scattering theory \cite{guh98} establishing a relation between the scattering matrix $\bS$ and the Green function $\bG$
\begin{equation} \label{eq:S0}
  \bS = \unity - \frac{2i\gamma\bG}{\unity+i\gamma\bG}\,,
\end{equation}
where $\gamma$ contains the information on the channel coupling, assumed to be the same for all channels, and $\bG$ is the
Green function.
In Hamiltonian systems $\bG$ is related to the system Hamiltonian $H$ via
\begin{equation} \label{eq:Gnuc}
  \bG = \frac{\unity}{\omega\,\unity-H}\,.
\end{equation}
Equation~(\ref{eq:S0}) has been derived originally in nuclear physics \cite{guh98} and later extended to quantum dots \cite{bee97}.
But the equation holds also in graphs \cite{kot99a,hof21}, where in this case $\bG$ is obtained from the inverse of the graph secular matrix $h$ truncated to the rows and columns corresponding to the coupling nodes.
In the present case with only one pair of channels $\bS$ and $\bG$ are $2\times 2$ matrices, and $\bG$ is the upper left block of $\bi{h}^{-1}$,
\begin{equation} \label{eq:Gh}
  \bG = \left(\bi{h}^{-1}\right)_{11}
    ={\left(\begin{array}{cc}
      (h^{-1})_{11} & (h^{-1})_{1\bar{1}} \\
      (h^{-1})_{\bar{1}1} & (h^{-1})_{\bar{1}\bar{1}}
    \end{array}\right)}\,.
\end{equation}
In the system, where $h_0$ is diagonal, the scattering equation~(\ref{eq:scatt}) is transformed into
\begin{equation} \label{eq:scatttilde}
  \tilde{\bb} = \tilde{\bS} \tilde{\ba}\,,
\end{equation}
where for the upper block
\begin{equation} \label{eq:abtilde}
  \tilde{\ba}(\omega,t) = \frac{\varepsilon}{\sqrt{2}} \ba(\omega,t)\,,
  \quad 
  \tilde{\bb} = \frac{\varepsilon}{\sqrt{2}} \bb(\omega,t)\,,
\end{equation}
see Eq.~(\ref{eq:rot}), and
\begin{equation} \label{eq:Stilde}
  \tilde{\bS} = \unity -  \frac{2i\gamma\tilde{\bG}}{\unity+i\gamma\tilde{\bG}}\,.
\end{equation}

$\tilde{\bG}$ has poles at the positions of the Kramers doublets (see Appendix~\ref{app:scatt} for details).
Expanding $\tilde{\bG}$ in terms of partial fractions, and restricting the discussion to the neighborhood of just one Kramers doublet, Eq.~(\ref{eq:Stilde}) reduces to
\begin{equation} \label{eq:S1}
  \tilde{\bS} = \unity - \frac{2i\gamma'}{(\omega+i\gamma')\unity -\tilde{H}}\,,
\end{equation}
with $\gamma'=\gamma g_n/2$, where $g_n$ is the residuum of the selected Kramers doublet at the position $\omega_n$.
$H$ is given by
\begin{equation} \label{eq:HNMR0}
  \tilde{H} = \omega_n \unity-\frac{\omega_0}{2} \sigma_z - \frac{\omega_1}{2} \sigma_x\,,
\end{equation}
where $\omega_0\sim a$, and $\omega_1\sim b$.
Equation~(\ref{eq:S1}) holds in the limit, where the line widths are small compared to the mean level spacing.
The notation has been chosen to be in accordance with NMR practice, see Eq.~(\ref{eq:HNMR}).
We have thus established explicitly the equivalence of the splitting of a Kramers doublet by the two perturbations with the Zeeman splitting of a spin \onehalf in magnetic fields in $x$ and $z$ directions.

The appearance of the Pauli matrices in Eq.~(\ref{eq:HNMR0}) should not surprise:
Symplectic symmetry means that all matrix elements of the Hamiltonian must be quaternionic real \cite{meh91,haa18}.
Hence for any perturbation of the symplectic symmetry there must be quaternionic imaginary matrix elements.
But because of the relation $\vec{\tau}=\frac{1}{i}\vec{\sigma}$ between the spin matrices and the quaternions these perturbations must be linear combinations of the spin matrices.

\subsection{The time-dependent case}

\subsubsection{The transformation into the rotating frame}

Now we turn to the situation, where $b$ is time dependent
\begin{equation} \label{eq:b}
  b(t) = 2b\cos\omega_Rt\,.
\end{equation}
Guided by the approach applied in NMR we look for a spin rotation removing the time dependence in the wave equation~(\ref{eq:wave}) for $\varphi_D$, without introducing new time dependencies in the wave equation for $\varphi_P$, and the continuity equation~(\ref{eq:cont}).
This is achieved by means of the rotation
\begin{equation} \label{eq:PhiR}
  \bphi_{AR}(x,t) = \bi{R}_t \tilde{\bphi}_A(x,t)
\end{equation}
with
\begin{equation} \label{eq:rot1}
  \bi{R}_t = \left(\begin{array}{cc}
    e^{-i\frac{\omega_Rt}{2}\sigma_z} & \cdot \\
    \cdot & e^{-i\frac{\omega_Rt}{2}\sigma_z} \\
  \end{array}\right)\,.
\end{equation}
$\tsigmaP$ and $\tsigmaD$ are not changed by this transformation.
The continuity equation~(\ref{eq:cont}) thus holds also for the $\tilde{\bPsi}_{AR}$ in the rotating frame.
The wave equation~(\ref{eq:wave}) is transformed into
\begin{equation} \label{eq:waveR}
  \left\{\square_R\,\unit + 2\tilde{\bV}_{AR} \frac{\partial^2}{\partial x^2}\right\} \bphi_{AR}(x,t) = 0\,.
\end{equation}
The time derivative in the d'Alembert operator is transformed by the rotation into
\begin{equation} \label{eq:dtR}
  \left(\frac{\partial}{\partial t}\right)_R = \bi{R}_t \frac{\partial}{\partial t} \bi{R}_t^\dag\,\unit
  = \frac{\partial}{\partial t}\,\unit + \frac{i\omega_R}{2}\sigdt\,.
\end{equation}
$\tilde{\bV}_P=a\tsigmaa$ is not changed, $\tilde{\bV}_P=\bV_{PR}$, but
\begin{eqnarray} \label{eq:VDR}
 \bV_{DR}
 &=& 2b\cos\omega_Rt \, \bi{R}_t \tsigmab \bi{R}_t^\dag\nonumber\\
 &=& 2b\cos\omega_Rt \, \bi{R}_t \frac{1}{\sqrt{2}}
    \left(\begin{array}{cc}
      -\sigma_x & -\sigma_x \\
      -\sigma_x & \sigma_x \\
    \end{array}\right) \bi{R}_t^\dag\nonumber\\
  &=& 2b\cos\omega_Rt \frac{1}{\sqrt{2}}
    \left(\begin{array}{cc}
      -\sigma_{xR} & -\sigma_{xR} \\
      -\sigma_{xR} & \sigma_{xR} \\
  \end{array}\right)\,.\nonumber
\end{eqnarray}
With
\begin{eqnarray} \label{eq:rel}
  2\cos\alpha\, \sigma_{xR} 
  &=& 2\cos\alpha\, e^{-i\frac{\alpha}{2}\sigma_z}\sigma_x e^{+i\frac{\alpha}{2}\sigma_z} \nonumber\\
  &=& 2\cos\alpha\, e^{-i \alpha\sigma_z}\sigma_x \nonumber\\
  &=& (1+e^{-2i\alpha\sigma_z})\sigma_x\,,\quad \alpha=\omega_Rt
\end{eqnarray}
one arrives at
\begin{equation} \label{eq:VDR1}
  \bV_{DR} 
  = \frac{b}{\sqrt{2}}
    \left(\begin{array}{cc}
      -\sigma_x & -\sigma_x \\
      -\sigma_x & \sigma_x \\
    \end{array}\right)
  = b\tsigmaD\,,
\end{equation}
where the terms rotating with $2\omega_Rt$ have been discarded.
Except for this neglect all time dependencies have disappeared in the rotation frame.
The $\tilde{\bV}_A$ are exactly the same as for the static case, there is only an additional term resulting from the transformation of the d'Alembert operator.
The situation is completely analogous to the one we met when transforming away the time dependence of the NMR Hamiltonian~(\ref{eq:HNMR}), see section~\ref{sec:intro}.

\subsubsection{The scattering matrix in the rotating frame}

By the transformation into the rotating frame the wave equation (\ref{eq:waveR}) has become time-independent.
We now can repeat the calculation for the time-independent case step by step.
First, the time dependence of the wave functions is separated,
\begin{equation} \label{eq:tsep1}
  \bphi_{AR}(x,t)=e^{-i\omega t}\bphi_{AR}(x)\,.
\end{equation}
By this separation the time derivative of the d'Alembert operator in the rotating frame~(\ref{eq:dtR}) is turned into $-i\left[\omega\unit-\frac{\omega_R}{2}\text{\tiny $\sigdt$}\right]$.
The equations obtained for the static case can thus be directly be adopted for the time-dependent case, if only $\omega$ is replaced by $\omega-\frac{\omega_R}{2}\sigma_z$.

Next, since in the rotating frame all time dependencies have disappeared, ordinary time-independent scattering theory can be applied again.
Equations~(\ref{eq:scatt}) and (\ref{eq:S0}) can be immediately transferred into the rotating frame,
\begin{equation} \label{eq:scattR}
  \bb_R = \bS_R\ba_R\,,
\end{equation}
where $\bS_R=\bi{R}_t\tilde{\bS} \bi{R}_t^\dag$.
The vectors of incoming and outgoing amplitudes in the rotating frame $\ba_R(\omega,t)=e^{-i\omega t}\ba_R(\omega)$ and $\bb_R(\omega,t)=e^{-i\omega t}\bb_R(\omega)$ are related to the corresponding quantities in the laboratory frame, $\tilde{\ba}(\omega,t)$ and $\tilde{\bb}(\omega,t)$, see Eq.~(\ref{eq:abtilde}), via spinor rotations,
\begin{equation} \label{eq:aR}
  \tilde{\ba}(\omega,t) 
    = e^{i\frac{\omega_Rt}{2}\sigma_z} \ba_R(\omega,t)
    = \left(\begin{array}{c}
      e^{-i\,\omega_-t}\,(\ba_R)_1(\omega) \\
      e^{-i\,\omega_+t}\,(\ba_R)_{\bar{1}}(\omega)\\
      \end{array}\right)\,,
\end{equation}
with $\omega_{\pm}=\omega\pm\frac{\omega_R}{2}$, see Eq.~(\ref{eq:rot1}).
A corresponding formula holds for $\tilde{\bb}(\omega,t)$.
The scattering matrix in the rotating frame is given by
\begin{equation} \label{eq:S1R}
  \bS_R(\omega) = \unity-\frac{2i\gamma'}{(\omega+i\gamma')\unity -H_R}\,,
\end{equation}
with
\begin{equation} \label{eq:HNMRR1}
 H_R = \omega_n\unity-\frac{\omega_0-\omega_R}{2}\sigma_z -\frac{\omega_1}{2}\sigma_x\,.
\end{equation}
$\bS_R(\omega)$ depends also on $\omega_R$ but this dependency has not been explicitly denoted to simplify the notation.
We have used that Eqs.~(\ref{eq:S1}) and (\ref{eq:HNMR0}) for the static case can be directly adopted for the time-dependent case by only replacing $\omega$ by $\omega-\frac{\omega_R}{2}\sigma_z$, see above.
We have obtained exactly the NMR Hamiltonian in the rotating frame, see Eq.~(\ref{eq:HNMRR}).

\section{Results}
\label{sec:results}

\subsection{From the laboratory to the rotating frame}

Our aim is to determine the spectral properties of the graph in the rotating frame meaning a determination of $\bS_R(\omega)$, see Eq.~(\ref{eq:S1R}).
Equation~(\ref{eq:aR}) shows that to this end up and down component in the tilted frame have to be excited at two different frequencies.
We only had one VNA at our disposal, so this could not be done, but there is an alternative:
Expression~(\ref{eq:aR}) shows that the upper left and lower down matrix elements of $\bS_R(\omega)$ may be interpreted as
\begin{equation} \label{eq:SR}
 (\bS_R)_{11}(\omega)=\tilde{S}_{11}(\omega_-)\,,\quad
 (\bS_R)_{\bar{1}\bar{1}}(\omega)=\tilde{S}_{\bar{1}\bar{1}}(\omega_+)\,.
\end{equation}
The diagonal elements of the scattering matrix in the rotating frame can thus be obtained from the ordinary scattering matrix elements taken at shifted frequencies.
The off-diagonal elements of $\bS_R$ cannot be evaluated in a similar way, as in this case an excitation of the graph at the frequency $\omega_+$, and a detection at $\omega_-$, and vice versa, is needed.
In principle this is experimentally realizable, but not with the existing equipment.
Fortunately for the present purpose knowledge of the trace of $\bS_R$, 
\begin{equation} \label{eq:TrSR}
 \mathrm{Tr}\bS_R(\omega)=\tilde{S}_{11}(\omega_-)+\tilde{S}_{\bar{1}\bar{1}}(\omega_+)
\end{equation}
is sufficient.

\begin{figure}
  \includegraphics[width=\linewidth]{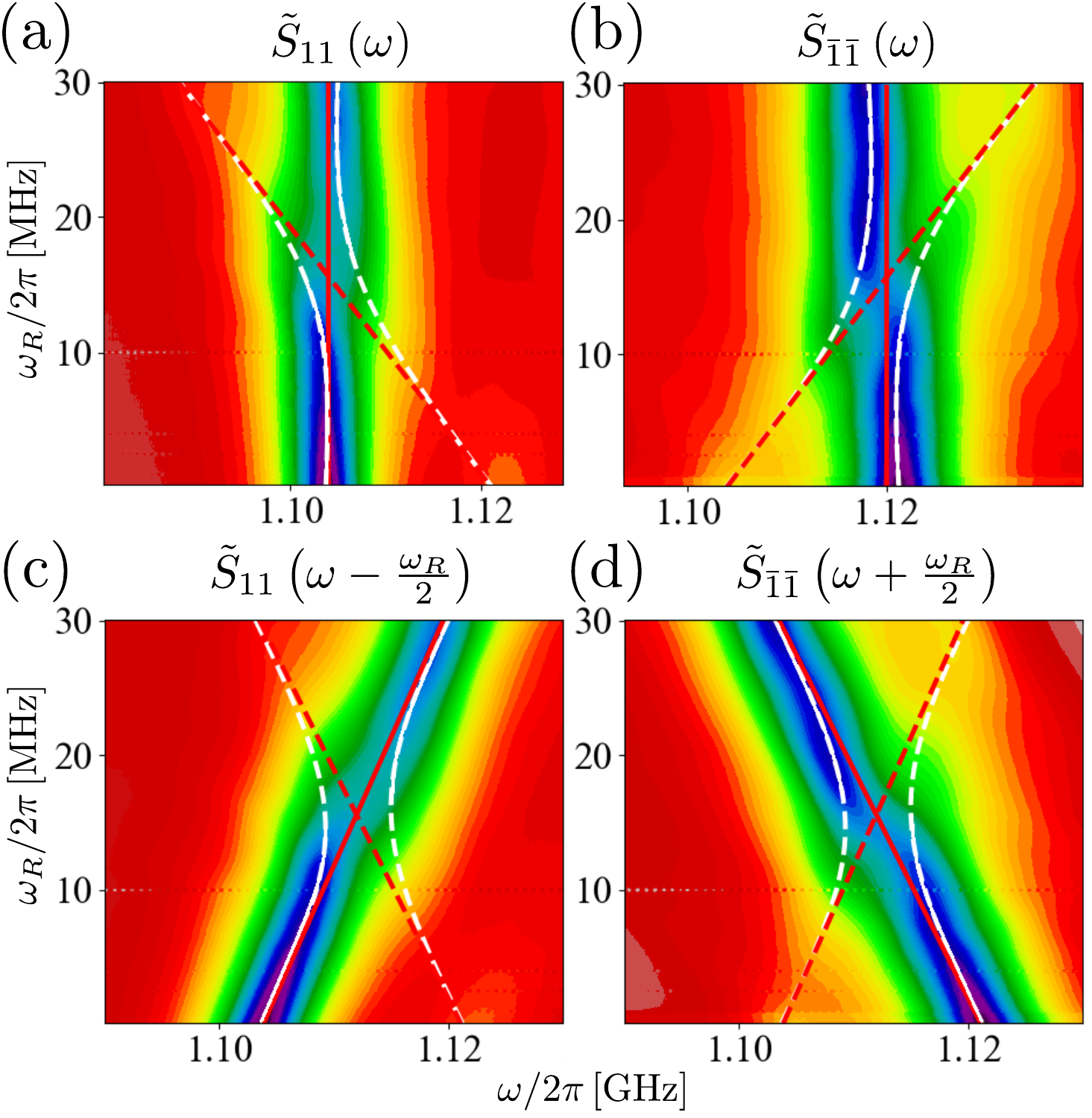}\\
  \caption{ \label{fig:labrot}
    Upper row:
    Reflection spectra $\tilde{S}_{11}(\omega)$ (a) and $\tilde{S}_{\bar{1}\bar{1}}(\omega)$ (b) as a function of frequency $\nu=\omega/2\pi$ for different radiofrequency frequencies $\nu_R=\omega_R/2\pi$.
    Lower row:
    $(\bS_R)_{11}(\omega)=\tilde{S}_{11}(\omega_-)$ (c) and $(\bS_R)_{\bar{1}\bar{1}}(\omega)=\tilde{S}_{\bar{1}\bar{1}}(\omega_+)$ (d), obtained by shear operations from the upper ones, see Eq.~(\ref{eq:SR}) and text.
  }
\end{figure}

The poles of of the scattering matrix $\bS_R(\omega)$ (\ref{eq:S1R}) in the rotating system are given, up to the absorption term, by the eigenvalues of $H_R$, see Eq.~(\ref{eq:eigenR}).
For $\omega_1=0$ the scattering matrix $\bS_R(\omega)$ is diagonal [see Eq.~(\ref{eq:S1R})] with diagonal elements
\begin{eqnarray} \label{eq:SR11} 
  (\bS_R)_{11}(\omega) &=& 1-\frac{2i\gamma'}{\omega_-+i\gamma'+\frac{\omega_0}{2}}=\tilde{S}_{11}(\omega_-)\,, \nonumber\\
  (\bS_R)_{\bar{1}\bar{1}}(\omega) &=& 1-\frac{2i\gamma'}{\omega_++i\gamma'-\frac{\omega_0}{2}}=\tilde{S}_{\bar{1}\bar{1}}(\omega_+)\,,
\end{eqnarray}
illustrating relations (\ref{eq:SR}) explicitly for the special case $\omega_1=0$.

\begin{figure}
  \includegraphics[width=\linewidth]{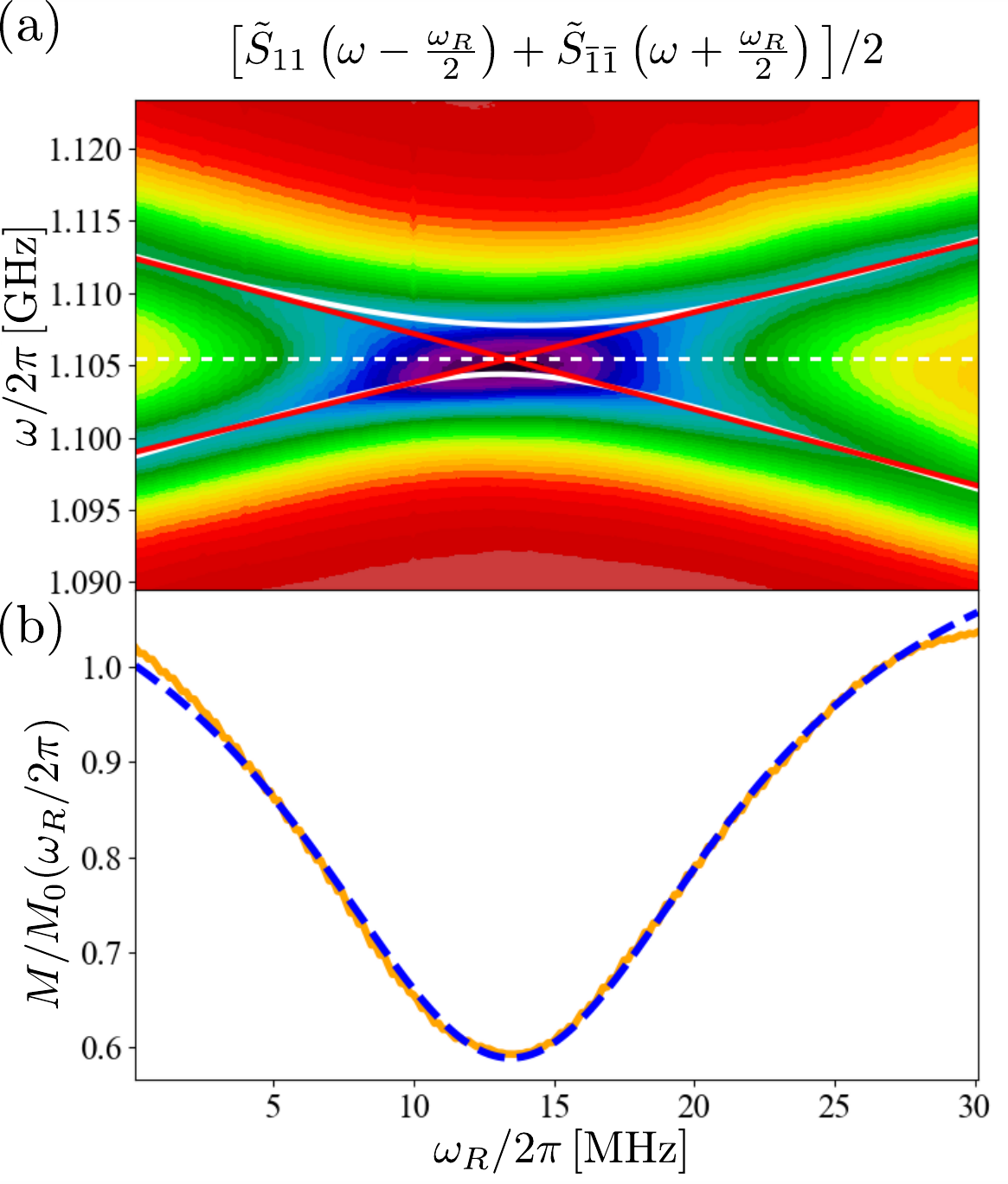}
  \caption{ \label{fig:Srot}
    (a) $\mathrm{Tr}\bS_R(\omega)$ obtained by adding $\tilde{S}_{11}(\omega_-)$ and $\tilde{S}_{\bar{1}\bar{1}}(\omega_+)$, shown in Figs.~\ref{fig:labrot}(c) and (d), and interchanging the axes.
    The hyperbolas plotted in white correspond to the eigenvalues of the Kramers doublet at 1.112\,GHz in the rotating frame.
    (b) Corresponding ``NMR'' resonance curve, see Eq.~(\ref{eq:res1}).
    The orange line has been extracted from the value of $\mathrm{Tr}\bS_R(\omega)$ along the dotted white symmetry line, the dashed blue line reflects the best fit with a Lorentzian.
  }
\end{figure}

$\tilde{S}_{11}(\omega)$ thus sees only the eigenvalue at $-\frac{\omega_0}{2}$, the lower frequency component of the Kramers doublet, and is blind for the other eigenvalue at $+\frac{\omega_0}{2}$.
For $\tilde{S}_{\bar{1}\bar{1}}(\omega)$ it is vice versa.
This is illustrated in Fig.~\ref{fig:labrot} for the Kramers doublet at 1.112\,GHz, Zeeman split into its components with frequencies at $\nu_1=\omega_1/2\pi= 1.105$\,GHz and $\nu_2=\omega_2/2\pi= 1.120$\,GHz, see the enlarged part of Fig.~\ref{fig:pilines}.
In Fig.~\ref{fig:labrot}(a) the eigenvalues of $\tilde{S}_{11}(\omega)$ for $\omega_1=0$ are plotted in red with $\omega_R$ versus $\omega$, the visible one with a solid line, the invisible one dashed.
Figure~\ref{fig:labrot}(b) shows the same for $\tilde{S}_{\bar{1}\bar{1}}(\omega)$.
In addition the eigenvalues are plotted in white for a small $\omega_1\ne 0$.
Now the straight lines found for $\omega_1=0$ are converted into hyperbolas.
Close to the avoided crossings both eigenvalues become visible both for $\tilde{S}_{11}(\omega)$ and $\tilde{S}_{\bar{1}\bar{1}}(\omega)$, but off the avoided crossings still only the vertical branches of the hyperbolas are visible.

In the same figures the experimental results are plotted.
The blue colors reflect the positions of the eigenvalues.
One finds exactly the predicted behavior.
For $\tilde{S}_{11}(\omega)$ there is a resonance at the lower Kramers component at $-\frac{\omega_0}{2}$, more or less independent of $\omega_R$.
Only for $\nu_R=\omega_R/2\pi= 15$\, MHz, corresponding to the splitting $\Delta\omega$ of the two Kramers doublets, there is an indication that something is happening.
The other Kramers component is not seen.
Exactly the complementary phenomenology is found for $\tilde{S}_{\bar{1}\bar{1}}(\omega)$.

These measurements are now converted into the rotating frame.
To this end we convert $\tilde{S}_{11}(\omega)$ and $\tilde{S}_{\bar{1}\bar{1}}(\omega)$ into $\tilde{S}_{11}(\omega_-)=(\bS_R)_{11}(\omega)$ and $\tilde{S}_{\bar{1}\bar{1}}(\omega_+)=(\bS_R)_{\bar{1}\bar{1}}(\omega)$, respectively.
This means a shear of the figures shown in the upper row, either to the left for $\tilde{S}_{11}(\omega)$, or to the right for $\tilde{S}_{\bar{1}\bar{1}}(\omega)$.
The result is shown in Figs.~\ref{fig:labrot}(c) and (d).
Due to the shear operations the hyperbolas now are the same in both figures.
In the last step we add up $\tilde{S}_{11}(\omega_-)$ and $\tilde{S}_{\bar{1}\bar{1}}(\omega_+)$ to obtain $\mathrm{Tr}\bS_R(\omega)$.
The result is shown in Fig.~\ref{fig:Srot}(a), now with $\omega$ and $\omega_R$ axes interchanged, to be in accordance with NMR conventions, see Fig.~\ref{fig:NMR}(b).

\begin{figure}
  \includegraphics[width=\linewidth]{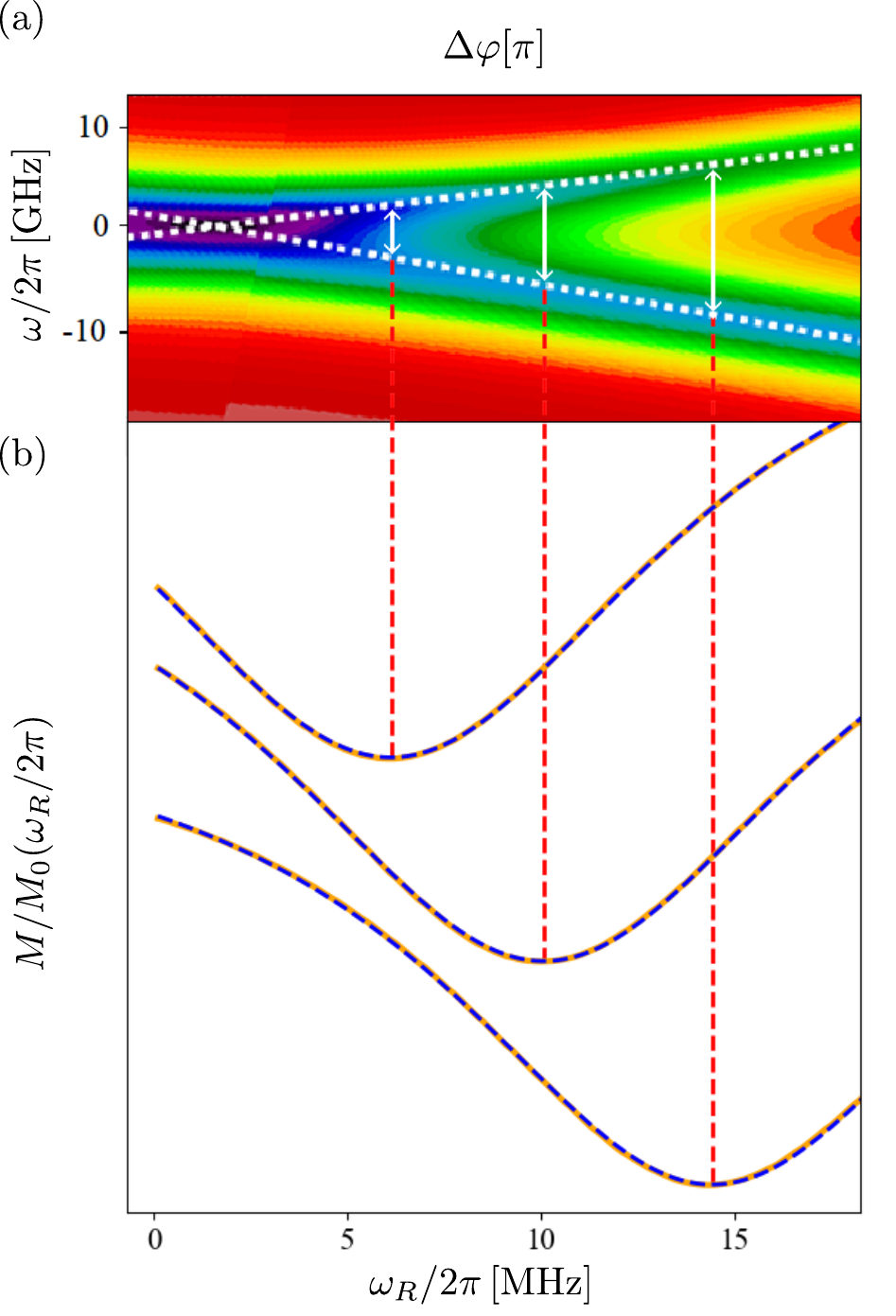}\\
  \caption{ \label{fig:lorenz}
    (a) Splitting of the Kramers doublet at 1.12\,GHz in dependence of $\omega_R$.
    The vertical red arrows mark the positions, where resonance experiments have been performed.
    (b) Corresponding resonance curves (orange lines) and Lorentzian fits (dashed blue lines), see Fig. \ref{fig:Srot}.
  }
\end{figure}

By means of this somewhat tricky operation we have been able to convert the measured $S$ matrix components $\tilde{S}_{11}(\omega)$ and $\tilde{S}_{\bar{1}\bar{1}}(\omega)$ in the laboratory frame into the $S$ matrix $\bS_R(\omega)$ in the rotating frame, with the small drawback that only the trace of this matrix is accessible with the existing set-up.
As a result we have obtained the spectrum of the NMR Hamiltonian in the rotating frame, a quantity not even available in a standard NMR experiment.

\begin{figure}
  \includegraphics[width=\linewidth]{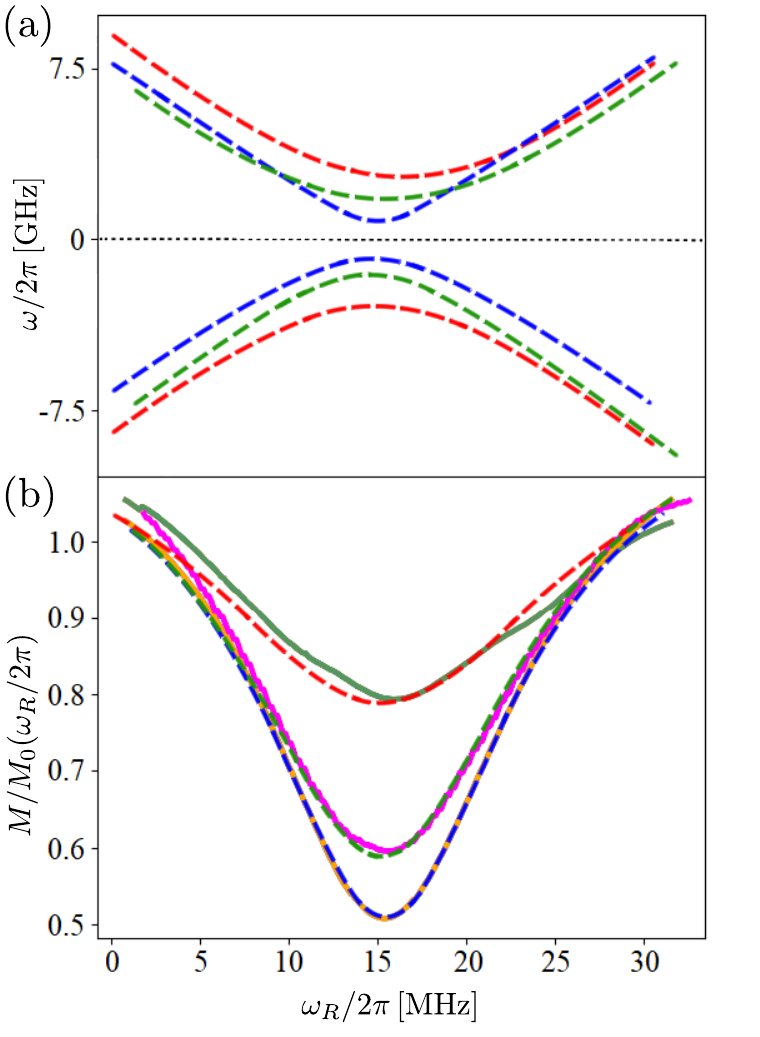}\\
  \caption{ \label{fig:omega1}
    (a) Eigenvalues of the Kramers doublet at 1.12\,GHz in the rotation frame for three different ``radio frequency'' strengths $\omega_1$, achieved by varying the length hubs induced by the diode.
    The dashed lines have been extracted from $\mathrm{Tr}\bS_R(\omega_R)$ by a fit with a pair of hyperbolas, see Fig.~\ref{fig:Srot}.
    (b) Corresponding resonance curves.
    The solid lines have been extracted from $\mathrm{Tr}\bS_R(0)$, see Fig.~\ref{fig:Srot}, the dashed lines reflect the best fits with a Lorentzian.
  }
\end{figure}

\subsection{The microwave analog of spin resonance}

Of particular interest is the behavior of $\mathrm{Tr}\bS_R(\omega)$ along the horizontal symmetry line corresponding to $\omega=0$, dotted in white in Fig.~\ref{fig:Srot}(a).
For this quantity one obtains from Eq.~(\ref{eq:S1R})
\begin{equation} \label{eq:res1}
  M(\omega_R) 
  = \frac{1}{2} \mathrm{Tr}\bS_R(0) 
  = 1-\frac{4\gamma'^2}{(\omega_0-\omega_R)^2 + \omega_1^2+4\gamma'^2}\,.
\end{equation}
This is exactly the formula for a typical Lorentzian shaped magnetic resonance curve, see Eq.~(\ref{eq:res}), with the only difference that there is an additional contribution to the resonance width from the coupling.
Figure~\ref{fig:Srot}(b) shows $M(\omega_R)$ as obtained from the value of $\mathrm{Tr}\bS_R(\omega)$ taken along the horizontal dashed white line in Fig.~\ref{fig:Srot}(a) and corresponding to $\mathrm{Tr}\bS_R(0)$.
There is a one-to-one correspondence between Figs.~\ref{fig:NMR}(b) and (c) from NMR with Figs.~\ref{fig:Srot}(a) and (b) from the microwave analog.

Two dependencies can be studied.
Equation~(\ref{eq:res1}) predicts a resonance for $\omega_R=\omega_0$.
This is illustrated in Fig.~\ref{fig:lorenz}.
The upper part of the figure shows the splitting of the Kramers doublet at 1.12\,GHz in dependence of $\omega_0$.
The axes have been interchanged in contrast to Fig.~\ref{fig:pilines}.
For the $\omega_0$ values marked by vertical lines a ``NMR'' has been performed.
The lower part of the figure shows the obtained resonance curves.
All resonance curves meet the resonance condition, and all of them are perfectly Lorentzian shaped.

Equation~(\ref{eq:res1}) predicts a resonance width of $\Delta\omega=\sqrt{\omega_1^2+4\gamma'^2}$ and a splitting of $2\Delta\omega$ of the two branches of the hyperbola in the rotating frame at the resonance position $\omega_R=\omega_0$.
This is illustrated in Fig.~\ref{fig:omega1}.
The upper part of the figure shows the eigenvalues in the rotating field for the Kramers doublet at 1.12\,GHz for three different $\omega_1$ values achieved by varying the length hubs induced by the diode in the dangling bonds, see appendix \ref{app:diode}.
The observed crossing of the hyperbolas should not occur, it reflects the experimental imperfections.
The lower part of the figures shows the corresponding ``NMR'' curves.
Again the line shapes are perfectly Lorentzian, and the line widths increase with $\omega_1$ as expected.
Further contributions to the line width are caused by absorption of the microwaves in the bonds, which in principle could be accounted for in terms of an imaginary contribution to $\omega$.
For a quantitative analysis of the resonance widths the data set was not sufficient.

\section{Conclusions}

We presented a completely new, so far unknown, approach towards spin resonance where no spins are involved.
It is based on the fact that a spin \onehalf is not really needed to study spin physics, a system exhibiting symplectic symmetry is sufficient.
A prerequisite for this achievement was the development of a technique allowing for rapid changes of the transmission properties of a network with frequencies up to 125 MHz.
Among other things this opens a new approach to spin relaxation studies.
In standard NMR relaxation measurements are performed to get information on the origin of the fluctuating interactions.
Here it is vice versa:
It is easy to generate well-controlled fluctuating interactions, again by means of diodes, and to study their implications for the resonance line shape.
In particular the old Bloembergen-Purcell-Pound model \cite{blo48} describing nuclear magnetic relaxation due to stochastic spin-flips could be directly addressed.

Furthermore, the here developed technique to change effective bond lengths by means of diodes enables the experimental access to a completely new class of systems inaccessible before, time dependent graphs, among others all types of Floquet systems.

The graph studied in the present work, in the absence of perturbations just a square with dangling bonds at the corners, does exhibit both symplectic and chiral symmetry.
All eigenvalues thus appear not only as Kramers doublets but also in pairs $\pm \omega_n$, see Eq.~(\ref{eq:zeros}).
This feature had been used by us previously for the study of the chiral ensembles in a square arrangement of coupled dielectric resonators \cite{reh20}.
This allows an interpretation of the two Kramers doublets, corresponding to $+\omega_n$ and $-\omega_n$, in terms of electron and positron states, respectively.
In fact the graph secular matrix $\bh$ (\ref{eq:sec2}) may alternatively be written completely in terms of the $\gamma$ matrices, and the column vectors $\bphi_A$ appearing as solutions of the wave equation~(\ref{eq:wave1}) may be interpreted as Dirac rank-4-spinors.
This aspect has not been treated in this paper, but it shows that the square graph is also an excellent candidate for the study of the complete four-dimensional Dirac equation, not only of its reduced two-dimensional version.

\appendix

\section{The diode}
\label{app:diode}

\begin{figure}
  \includegraphics[width=\linewidth]{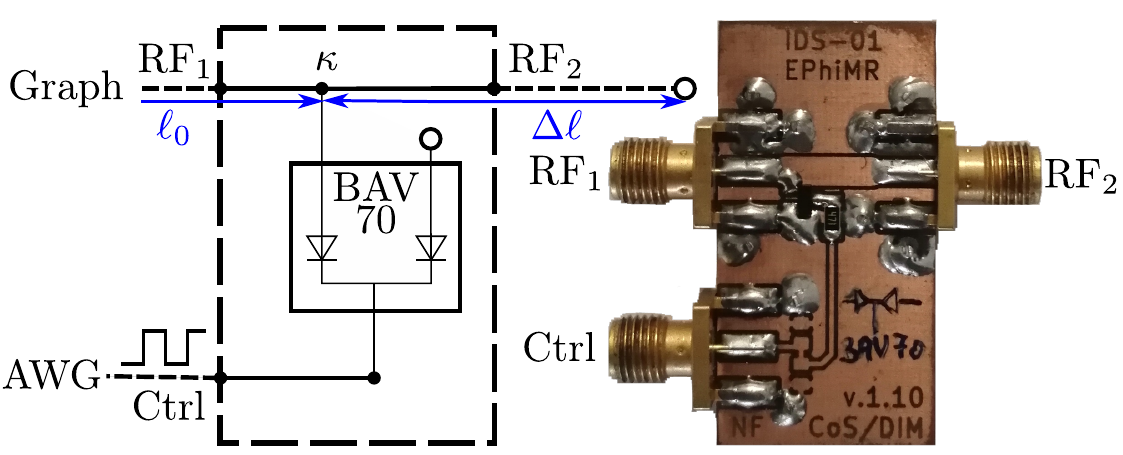}
  \caption{ \label{fig:diode}
    Sketch and photograph of the diode construction attached at the end of the dangling bond denoted by D in Fig.~\ref{fig:sketch}(b).
    Depending on the state of the control signal (Ctrl), generated by a waveform generator (AWG), the microwaves entering via port RF$_1$ are either reflected at $\kappa$, the connection point to the diode BAV70, (high state) or after traveling through an additional bond of length of $\Delta\ell$ attached at port RF$_2$ (low state).
  }
\end{figure}

Figure~\ref{fig:diode} shows sketch and photograph of the diode construction attached at the end of the dangling bond denoted by D in Fig.~\ref{fig:sketch}(b).
The high speed double diode BAV~70 allows for bipolar flips with frequencies up to $125$\,MHz.
At the other dangling bond, denoted in Fig.~\ref{fig:sketch}(b) by D$^\prime$, a double diode BAV~56 with opposite polarity is attached.
Thus the lengths $l_D$ of the dangling bonds can be periodically switched anti-cyclically between $l_0$ and $l_0+\Delta \ell$, and $l_0+\Delta \ell$ and $l_0$, respectively, allowing for a switching without changing the total length.

The wavefunctions meeting at the T junction may be written as
\begin{equation}\label{eq:psixL}
  \psi_l(x_l)=a_le^{-ikx_l}-b_le^{ikx_l}\,,\quad l=0,1,2\,,
\end{equation}
where the $x_l$ are the distances from the T junction.
The vectors $a=(a_0,a_1,a_2)^T$ and $b=(b_0,b_1,b_2)^T$ of the amplitudes of the in- and outgoing waves obey the scattering equation
\begin{equation}\label{eq:scatta}
  b=S_Ta\,,
\end{equation}
where
\begin{equation}\label{eq:ST}
  S_T=\frac{1}{3}\left(\begin{array}{rrr}
       1 & -2 & -2 \\
      -2 &  1 & -2 \\
      -2 & -2 &  1
    \end{array}\right)
\end{equation}
is the scattering matrix of the T junction.
Assuming that the dangling bond is attached to port 0 and short terminated at the end, the amplitude of the incoming wave is obtained from the amplitude of the outgoing wave just by multiplication with a phase factor,
\begin{equation}\label{eq:phase}
  a_0=e^{2ikl_D} b_0\,,
\end{equation}
where $l_D$ is the length of the bond.
Eliminating $a_0, b_0$ from the scattering equation system (\ref{eq:scatta}) using Eq.~(\ref{eq:phase}), a $2\times2$ scattering matrix equation is obtained for the amplitudes of the in- and outgoing waves in bonds 1 and 2,
\begin{equation}\label{eq:scattb}
  \left(\begin{array}{c}
    b_1 \\
    b_2 \\
  \end{array}\right)
  = S_V \left(\begin{array}{c}
    a_1 \\
    a_2 \\
  \end{array}\right)\,,
\end{equation}
where
\begin{equation}\label{eq:SV}
  S_V = \frac{1}{\cos\alpha-2i\sin\alpha}
    \left(\begin{array}{cc}
        \cos\alpha & 2i\sin\alpha \\
      2i\sin\alpha &   \cos\alpha \\
    \end{array}\right)\,,
\end{equation}
with $\alpha = kl_D$.
Introducing another angle $\beta$ via
\begin{equation}\label{eq:beta}
  \tan\beta=2\tan\alpha\,.
\end{equation}
Equation~(\ref{eq:SV}) may be alternatively written as
\begin{equation}\label{eq:SV1}
  S_V=e^{i\beta}\left(
    \begin{array}{cc}
       \cos\beta & i\sin\beta \\
      i\sin\beta &  \cos\beta \\
    \end{array}\right)\,.
\end{equation}
In its diagonal basis the scattering matrix is given by
\begin{equation}\label{eq:SV2}
    (S_V)_D=\mathrm{diag}(e^{2i\beta},1)
\end{equation}
Expanding $l_D=l_0+\Delta l$ in first order in $\Delta l$, one obtains using Eq.~(\ref{eq:beta}),
\begin{equation}\label{eq:SV3}
   (S_V)_D= (S_V)_{D0}\,
    \mathrm{diag}(e^{ik(\Delta l)_\mathrm{eff}},1)\,,
\end{equation}
where $(S_V)_{D0}$ is the value of $(S_V)_D$ for $l_D=l_0$, and
\begin{equation}\label{eq:leff}
  (\Delta l)_\mathrm{eff}= \frac{4\Delta l}{1+3\sin^2(kl_0)}\,.
\end{equation}
Equation~(\ref{eq:SV3}) shows that part of the scattering matrix is not influenced by the dangling bond.
But for the other part a change of the length of the dangling bond by $\Delta l$ produces a phase factor equivalent to a change of the length of the sum of the two bonds meeting at the T junction by $(\Delta l)_\mathrm{eff}$.

\section{Useful matrix transformations}
\label{app:trans}
\begin{table}
  \caption{\label{tab:trafos}
    Matrix transformations into the basis of $\bh_0$
  }
\begin{ruledtabular}
\begin{tabular}{|c|c|r|}
\hline
 & $\bi{A}$ & \hfill$\tilde{\bi{A}}$ \hspace*{\fill} \\
 \hline & & \\[-2.5ex]
 $ \sigmaP$ & $\matfix{\cdot}{i\sigma_y}{-i\sigma_y}{\cdot}$
 & $\frac{\displaystyle 1}{\displaystyle\sqrt{2}} \matfix{\unity}{-\unity}{-\unity}{-\unity}$ \\[3ex]
 \hline & & \\[-2.5ex]
 $ \sigmaD$ & $\matfix{\cdot}{\unity}{\unity}{\cdot}$
 & $\frac{\displaystyle 1}{\displaystyle\sqrt{2}}\matfix{\unity}{\unity}{\unity}{-\unity}$ \\[3ex]
 \hline & & \\[-2.5ex]
 $\sigmaa$ & $\matfix{\sigma_z}{\cdot}{\cdot}{-\sigma_z}$
 & $\frac{\displaystyle 1}{\displaystyle\sqrt{2}} \matfix{\sigma_z}{-\sigma_z}{-\sigma_z}{-\sigma_z}$ \\[3ex]
 \hline & & \\[-2.5ex]
 $ \sigmab$ & $\matfix{\sigma_z}{\cdot}{\cdot}{\sigma_z}$
 & $\frac{\displaystyle 1}{\displaystyle\sqrt{2}} \matfix{-\sigma_x}{-\sigma_x}{-\sigma_x}{\sigma_x}$ \\[3ex]
 \hline & & \\[-2.5ex]
$ \sigmaP\sigmaa$ & $\matfix{\cdot}{\sigma_x}{\sigma_x}{\cdot}$
 & $\matfix{\sigma_z}{\cdot}{\cdot}{\sigma_z}$ \\[3ex]
 \hline & & \\[-2.5ex]
 $ \sigmaD\sigmab$ & $\matfix{\cdot}{\sigma_z}{\sigma_z}{\cdot}$
 & $\matfix{-\sigma_x}{\cdot}{\cdot}{-\sigma_x}$ \\[3ex]
 \hline
\end{tabular}
\end{ruledtabular}
\end{table}
For the readers convenience in Table~\ref{tab:trafos} all matrix transformations into the basis system of $\bh_0$ are collected which are needed, see Eq.~(\ref{eq:trans}).

\section{The secular matrix}
\label{app:sec}

After separation of the time dependence the time dependent wave equation~(\ref{eq:wave}) turns into
\begin{equation} \label{eq:wave2}
  \left\{\left[k^2+ \frac{d^2}{dx^2}\right]\unit+2\bV_A\frac{\partial^2}{\partial x^2}\right\}
  \bphi_A(x) = 0
\end{equation}
($A=B$, $P$, $D$), with the $\bV_A$ given by Eq.~(\ref{eq:VA}).
It follows
\begin{equation} \label{eq:wave3}
  \bphi_A''(x) = -\bK_A^2 \bphi_A(x)\,,
\end{equation}
where
\begin{equation} \label{eq:KA}
 \bK_A = k\left[\unit-\bV_A\right]
\end{equation}
(correct in linear order of the perturbations).
The equation is solved immediately with the result
\begin{equation} \label{eq:sol1}
  \bphi_A(x) = \frac{\sin \bK_A(l-x)}{\sin \bK_Al}\bphi_0\,,\quad \bphi_0
  = \left(\begin{array}{c}
    \varphi_1 \\
    \varphi_{\bar{1}} \\
    \varphi_{2} \\
    \varphi_{\bar{2}} \\
  \end{array}\right)\,,
\end{equation}
where the boundary conditions~(\ref{eq:phi}) have been taken into account.
Plugging in the result into the continuity equation~(\ref{eq:cont}) one obtains for the secular matrix
\begin{equation} \label{eq:sec1}
  \bh = -\bF_B - \bF_P - \bF_D + \sigmaP \biG_P + \sigmaD \biG_D\,,
\end{equation}
where
\begin{equation} \label{eq:FG}
  \bF_A = \bK_A \cot(\bK_Al)\,,\quad \biG_A = \bK_A/\sin(\bK_Al)\,.
\end{equation}
Expansion of the $\bF_A$, $\biG_A$ in linear order in the perturbations $a$, $b$ yields
\begin{equation} \label{eq:fexp}
  \bF_A = f\unit - k f'\bV_A\,,\quad \biG_A = g\unit - kg'\bV_A\,.
\end{equation}
One thus obtains
\begin{equation} \label{eq:sec3}
  \bh = \bh_0 + \bh_a + \bh_b
\end{equation}
with $\bh_0$ given by Eq.~(\ref{eq:h0}), and
\begin{eqnarray} \label{eq:hab1}
  \bh_a &=& kf'\bV_P-kg'\sigmaP \bV_P\nonumber\\
    &=& ak\left[f'\sigmaa-g'\sigmaP\sigmaa\right] \nonumber\\[1ex]
  \bh_b &=& kf'\bV_D-kg'\sigmaD \bV_D\nonumber\\
    &=& bk\left[f'\sigmab-g'\sigmaD\sigmab\right]
\end{eqnarray}
After transformation into the basis, where $\bh_0$ is diagonal (see table~\ref{tab:trafos}), Eqs.~(\ref{eq:hab}) are obtained.

\section{The scattering matrix}
\label{app:scatt}

Using the transformation~(\ref{eq:rot}), $\bG=(\bi{h}^{-1})_{11}$ may be written as
\begin{equation} \label{eq:G1}
  \bG = \tilde{\bi{r}}^\dag\tilde{\bi{h}}^{-1}\tilde{\bi{r}}\,,
  \quad
  \tilde{r} = \frac{1}{\sqrt{2}}
  \left(\begin{array}{c}
    \varepsilon_y \\
    i\sigma_y\varepsilon_y \\
  \end{array}\right)\,,
\end{equation}
where $\tilde{\bi{r}}$ is the left column of $\tilde{\bi{R}}$, or
\begin{equation} \label{eq:G2}
  \bG = \frac{1}{2} \varepsilon_y^\dag \left(\tilde{h}_+^{-1}
      + \sigma_y\tilde{h}_-^{-1} \sigma_y\right) \varepsilon_y
    = \varepsilon_y^\dag \tilde{\bG} \varepsilon_y
\end{equation}
with
\begin{eqnarray} \label{eq:G4}
  \tilde{\bG} &=& \frac{1}{2}\left(\tilde{h}_+^{-1}
    + \sigma_y \tilde{h}_-^{-1} \sigma_y \right) \nonumber\\
  &=& \frac{1}{2} \Big(\frac{\unity}{h_{0+} \unity + (a\sigma_z-b\sigma_x)k_+} \nonumber\\
    && \qquad + \sigma_y\frac{\unity}{h_{0-}\unity-(a\sigma_z-b\sigma_x)k_-}\sigma_y\Big) \nonumber\\
  &=&\frac{1}{2} \Big(\frac{\unity}{h_{0+}\unity\nonumber+(a\sigma_z-b\sigma_x)k_+}\nonumber\\
    && \qquad +\frac{\unity}{h_{0-}\unity + (a\sigma_z-b\sigma_x)k_-}\Big)
\end{eqnarray}
where
\begin{equation} \label{eq:hk}
 h_{0\pm}=-3f\pm g\sqrt{2}\,,\quad k_\pm=\left(\frac{f'}{\sqrt{2}}\mp g'\right)k\,,
\end{equation}
see Eqs.~(\ref{eq:h0}) and (\ref{eq:hab}).
In the last step of Eq.~(\ref{eq:G4}) it has been used that the spin matrices are anticommuting.

Plugging in this result into Eq.~(\ref{eq:S0}), we obtain
\begin{equation} \label{eq:Stilde1}
  \bS=\varepsilon_y^\dag\tilde{\bS}\varepsilon_y\,,
\end{equation}
where
\begin{equation} \label{eq:Stilde2}
  \tilde{\bS}= \unity-2i\gamma\tilde{\bG}\frac{\unity}{\unity+i\gamma\tilde{\bG}}
\end{equation}

From Eq.~(\ref{eq:G4}) an analytic expansion of $\tilde{\bG}$ in terms of partial fractions can be obtained.
This is illustrated here for $\tilde{h}_+^{-1}$ in the absence of the perturbation,
\begin{eqnarray} \label{eq:partfrac}
  \tilde{h}_{0+}^{-1} 
  &=& \frac{1}{k(-3\cot\varphi+\sqrt{2}/\sin\varphi)}\,,\quad \varphi=kl\nonumber\\
  &=&\frac{1}{3k}\frac{\sin\varphi}{\cos\varphi_0-\cos\varphi}\,,
\end{eqnarray}
where $\cos\varphi_0=\sqrt{2}/3$, see Eq.~(\ref{eq:zeros}).
The latter expression can be written as
\begin{equation} \label{eq:partfrac1}
  \tilde{h}_{0+}^{-1} = \frac{1}{6k}\left(\cot\varphi_++\cot\varphi_-\right)\,,\quad \varphi_\pm=\frac{\varphi\pm\varphi_0}{2}
\end{equation}
Using the well-known partial fraction expansion for the cotangent, the familiar pole expansion of $\tilde{\bG}$ is obtained,
\begin{equation} \label{eq:Gpole}
 \tilde{\bG} = \sum\limits_n \frac{g_n}{k-k_n}\,,
\end{equation}
where the sum is over all Kramers doublets.
The extension to the situation where there is a perturbation is straightforward.
We do not proceed further in this direction, the details are not of relevance in the present context.
Instead we expand $\tilde{h}_{0+}$ and $\tilde{h}_{0-}$ close to its zeros at $\varphi_{n\pm}=\pm\varphi_0+n\pi$, here as an example for the zero of $\tilde{h}_{0+}$ at $\varphi=\varphi_0$,

\begin{equation} \label{eq:hzero}
  \tilde{h}_{0+} =
  3k\frac{\cos\varphi_0-\cos\varphi}{\sin\varphi}=3k_0\left(\varphi-\varphi_0\right)+{\cal O} \left(\varphi-\varphi_0\right)^2
\end{equation}
with $k_0=\varphi_0/l$.
The same expression, up to possible minus signs, is obtained for all zeros.
Entering with expansion (\ref{eq:Gpole}) into Eq.~(\ref{eq:G4}), restricting the expansion to just one term, and renaming the variables appropriately, Eqs.~(\ref{eq:S1}) and (\ref{eq:HNMR0}) are obtained.

\begin{acknowledgments}
H.-J.~St. thanks the department of physics of the university of Marburg for providing him with every support needed to continue with his research over many years after his official retirement.
\end{acknowledgments}


\begin{thebibliography}{18}%
	\makeatletter
	\providecommand \@ifxundefined [1]{%
		\@ifx{#1\undefined}
	}%
	\providecommand \@ifnum [1]{%
		\ifnum #1\expandafter \@firstoftwo
		\else \expandafter \@secondoftwo
		\fi
	}%
	\providecommand \@ifx [1]{%
		\ifx #1\expandafter \@firstoftwo
		\else \expandafter \@secondoftwo
		\fi
	}%
	\providecommand \natexlab [1]{#1}%
	\providecommand \enquote  [1]{``#1''}%
	\providecommand \bibnamefont  [1]{#1}%
	\providecommand \bibfnamefont [1]{#1}%
	\providecommand \citenamefont [1]{#1}%
	\providecommand \href@noop [0]{\@secondoftwo}%
	\providecommand \href [0]{\begingroup \@sanitize@url \@href}%
	\providecommand \@href[1]{\@@startlink{#1}\@@href}%
	\providecommand \@@href[1]{\endgroup#1\@@endlink}%
	\providecommand \@sanitize@url [0]{\catcode `\\12\catcode `\$12\catcode
		`\&12\catcode `\#12\catcode `\^12\catcode `\_12\catcode `\%12\relax}%
	\providecommand \@@startlink[1]{}%
	\providecommand \@@endlink[0]{}%
	\providecommand \url  [0]{\begingroup\@sanitize@url \@url }%
	\providecommand \@url [1]{\endgroup\@href {#1}{\urlprefix }}%
	\providecommand \urlprefix  [0]{URL }%
	\providecommand \Eprint [0]{\href }%
	\providecommand \doibase [0]{https://doi.org/}%
	\providecommand \selectlanguage [0]{\@gobble}%
	\providecommand \bibinfo  [0]{\@secondoftwo}%
	\providecommand \bibfield  [0]{\@secondoftwo}%
	\providecommand \translation [1]{[#1]}%
	\providecommand \BibitemOpen [0]{}%
	\providecommand \bibitemStop [0]{}%
	\providecommand \bibitemNoStop [0]{.\EOS\space}%
	\providecommand \EOS [0]{\spacefactor3000\relax}%
	\providecommand \BibitemShut  [1]{\csname bibitem#1\endcsname}%
	\let\auto@bib@innerbib\@empty
	\bibitem [{\citenamefont {Mehta}(1991)}]{meh91}%
	\BibitemOpen
	\bibfield  {author} {\bibinfo {author} {\bibfnamefont {M.~L.}\ \bibnamefont
			{Mehta}},\ }\href {https://doi.org/10.1016/C2009-0-22297-5} {\emph {\bibinfo
			{title} {Random Matrices (Revised and Enlarged Second Edition)}}}\ (\bibinfo
	{publisher} {Academic Press},\ \bibinfo {address} {San Diego},\ \bibinfo
	{year} {1991})\BibitemShut {NoStop}%
	\bibitem [{\citenamefont {Haake}\ \emph {et~al.}(2018)\citenamefont {Haake},
		\citenamefont {Gnutzmann},\ and\ \citenamefont {Ku{\'s}}}]{haa18}%
	\BibitemOpen
	\bibfield  {author} {\bibinfo {author} {\bibfnamefont {F.}~\bibnamefont
			{Haake}}, \bibinfo {author} {\bibfnamefont {S.}~\bibnamefont {Gnutzmann}},\
		and\ \bibinfo {author} {\bibfnamefont {M.}~\bibnamefont {Ku{\'s}}},\ }\href
	{https://doi.org/10.1007/978-3-319-97580-1} {\emph {\bibinfo {title} {Quantum
				Signatures of Chaos. 4th edition}}}\ (\bibinfo  {publisher} {Springer},\
	\bibinfo {address} {Berlin},\ \bibinfo {year} {2018})\BibitemShut {NoStop}%
	\bibitem [{\citenamefont {St{\"o}ckmann}\ and\ \citenamefont
		{Kuhl}(2022)}]{stoe22}%
	\BibitemOpen
	\bibfield  {author} {\bibinfo {author} {\bibfnamefont {H.-J.}\ \bibnamefont
			{St{\"o}ckmann}}\ and\ \bibinfo {author} {\bibfnamefont {U.}~\bibnamefont
			{Kuhl}},\ }\bibfield  {title} {\bibinfo {title} {Microwave studies of the
			spectral statistics in chaotic systems},\ }\href
	{https://doi.org/10.1088/1751-8121/ac87e0} {\bibfield  {journal} {\bibinfo
			{journal} {J. Phys. A}\ }\textbf {\bibinfo {volume} {55}},\ \bibinfo {pages}
		{383001} (\bibinfo {year} {2022})}\BibitemShut {NoStop}%
	\bibitem [{\citenamefont {Joyner}\ \emph {et~al.}(2014)\citenamefont {Joyner},
		\citenamefont {M{\"u}ller},\ and\ \citenamefont {Sieber}}]{joy14}%
	\BibitemOpen
	\bibfield  {author} {\bibinfo {author} {\bibfnamefont {C.~H.}\ \bibnamefont
			{Joyner}}, \bibinfo {author} {\bibfnamefont {S.}~\bibnamefont {M{\"u}ller}},\
		and\ \bibinfo {author} {\bibfnamefont {M.}~\bibnamefont {Sieber}},\
	}\bibfield  {title} {\bibinfo {title} {{GSE} statistics without spin},\
	}\href {https://doi.org/10.1209/0295-5075/107/50004} {\bibfield  {journal}
		{\bibinfo  {journal} {Europhys. Lett.}\ }\textbf {\bibinfo {volume} {107}},\
		\bibinfo {pages} {50004} (\bibinfo {year} {2014})}\BibitemShut {NoStop}%
	\bibitem [{\citenamefont {Rehemanjiang}\ \emph {et~al.}(2016)\citenamefont
		{Rehemanjiang}, \citenamefont {Allgaier}, \citenamefont {Joyner},
		\citenamefont {M{\"u}ller}, \citenamefont {Sieber}, \citenamefont {Kuhl},\
		and\ \citenamefont {St{\"o}ckmann}}]{reh16}%
	\BibitemOpen
	\bibfield  {author} {\bibinfo {author} {\bibfnamefont {A.}~\bibnamefont
			{Rehemanjiang}}, \bibinfo {author} {\bibfnamefont {M.}~\bibnamefont
			{Allgaier}}, \bibinfo {author} {\bibfnamefont {C.~H.}\ \bibnamefont
			{Joyner}}, \bibinfo {author} {\bibfnamefont {S.}~\bibnamefont {M{\"u}ller}},
		\bibinfo {author} {\bibfnamefont {M.}~\bibnamefont {Sieber}}, \bibinfo
		{author} {\bibfnamefont {U.}~\bibnamefont {Kuhl}},\ and\ \bibinfo {author}
		{\bibfnamefont {H.-J.}\ \bibnamefont {St{\"o}ckmann}},\ }\bibfield  {title}
	{\bibinfo {title} {Microwave realization of the {G}aussian symplectic
			ensemble},\ }\href {https://doi.org/10.1103/PhysRevLett.117.064101}
	{\bibfield  {journal} {\bibinfo  {journal} {Phys. Rev. Lett.}\ }\textbf
		{\bibinfo {volume} {117}},\ \bibinfo {pages} {064101} (\bibinfo {year}
		{2016})}\BibitemShut {NoStop}%
	\bibitem [{\citenamefont {Rehemanjiang}\ \emph {et~al.}(2018)\citenamefont
		{Rehemanjiang}, \citenamefont {Richter}, \citenamefont {Kuhl},\ and\
		\citenamefont {St{\"o}ckmann}}]{reh18}%
	\BibitemOpen
	\bibfield  {author} {\bibinfo {author} {\bibfnamefont {A.}~\bibnamefont
			{Rehemanjiang}}, \bibinfo {author} {\bibfnamefont {M.}~\bibnamefont
			{Richter}}, \bibinfo {author} {\bibfnamefont {U.}~\bibnamefont {Kuhl}},\ and\
		\bibinfo {author} {\bibfnamefont {H.-J.}\ \bibnamefont {St{\"o}ckmann}},\
	}\bibfield  {title} {\bibinfo {title} {Spectra and spectral correlations of
			microwave graphs with symplectic symmetry},\ }\href
	{https://doi.org/10.1103/PhysRevE.97.022204} {\bibfield  {journal} {\bibinfo
			{journal} {Phys. Rev. E}\ }\textbf {\bibinfo {volume} {97}},\ \bibinfo
		{pages} {022204} (\bibinfo {year} {2018})}\BibitemShut {NoStop}%
	\bibitem [{\citenamefont {Lu}\ \emph {et~al.}(2020)\citenamefont {Lu},
		\citenamefont {Che}, \citenamefont {Zhang},\ and\ \citenamefont
		{Dietz}}]{lu20}%
	\BibitemOpen
	\bibfield  {author} {\bibinfo {author} {\bibfnamefont {J.}~\bibnamefont
			{Lu}}, \bibinfo {author} {\bibfnamefont {J.}~\bibnamefont {Che}}, \bibinfo
		{author} {\bibfnamefont {X.}~\bibnamefont {Zhang}},\ and\ \bibinfo {author}
		{\bibfnamefont {B.}~\bibnamefont {Dietz}},\ }\bibfield  {title} {\bibinfo
		{title} {Experimental and numerical investigation of parametric spectral
			properties of quantum graphs with unitary or symplectic symmetry},\ }\href
	{https://doi.org/10.1103/PhysRevE.102.022309} {\bibfield  {journal} {\bibinfo
			{journal} {Phys. Rev. E}\ }\textbf {\bibinfo {volume} {102}},\ \bibinfo
		{pages} {022309} (\bibinfo {year} {2020})}\BibitemShut {NoStop}%
	\bibitem [{\citenamefont {Ma}\ and\ \citenamefont {Anlage}(2023)}]{ma23}%
	\BibitemOpen
	\bibfield  {author} {\bibinfo {author} {\bibfnamefont {S.}~\bibnamefont
			{Ma}}\ and\ \bibinfo {author} {\bibfnamefont {S.~M.}\ \bibnamefont
			{Anlage}},\ }\bibfield  {title} {\bibinfo {title} {Experimental realization
			of anti-unitary wave-chaotic photonic topological insulator graphs showing
			kramers degeneracy and symplectic ensemble statistics},\ }\href
	{https://doi.org/https://doi.org/10.1002/adom.202301852} {\bibfield
		{journal} {\bibinfo  {journal} {Advanced Optical Materials}\ }\textbf
		{\bibinfo {volume} {n/a}},\ \bibinfo {pages} {2301852} (\bibinfo {year}
		{2023})}\BibitemShut {NoStop}%
	\bibitem [{\citenamefont {{\L}awniczak}\ \emph {et~al.}(2024)\citenamefont
		{{\L}awniczak}, \citenamefont {Akhshani}, \citenamefont {Farooq},
		\citenamefont {Bauch},\ and\ \citenamefont {Sirko}}]{law24}%
	\BibitemOpen
	\bibfield  {author} {\bibinfo {author} {\bibfnamefont {M.}~\bibnamefont
			{{\L}awniczak}}, \bibinfo {author} {\bibfnamefont {A.}~\bibnamefont
			{Akhshani}}, \bibinfo {author} {\bibfnamefont {O.}~\bibnamefont {Farooq}},
		\bibinfo {author} {\bibfnamefont {S.}~\bibnamefont {Bauch}},\ and\ \bibinfo
		{author} {\bibfnamefont {L.}~\bibnamefont {Sirko}},\ }\bibfield  {title}
	{\bibinfo {title} {Experimental distributions of the reflection amplitude for
			networks with unitary and symplectic symmetries},\ }\href
	{https://doi.org/10.12693/APhysPolA.144.469} {\bibfield  {journal} {\bibinfo
			{journal} {Acta Physica Polonica A}\ }\textbf {\bibinfo {volume} {144}},\
		\bibinfo {pages} {469} (\bibinfo {year} {2024})}\BibitemShut {NoStop}%
	\bibitem [{\citenamefont {Slichter}(1980)}]{sli80}%
	\BibitemOpen
	\bibfield  {author} {\bibinfo {author} {\bibfnamefont {C.~P.}\ \bibnamefont
			{Slichter}},\ }\href@noop {} {\emph {\bibinfo {title} {Principles of Magnetic
				Resonance}}}\ (\bibinfo  {publisher} {Springer},\ \bibinfo {address}
	{Berlin},\ \bibinfo {year} {1980})\BibitemShut {NoStop}%
	\bibitem [{\citenamefont {Hofmann}\ \emph {et~al.}(2024)\citenamefont
		{Hofmann}, \citenamefont {Schmidt}, \citenamefont {St\"ockmann},\ and\
		\citenamefont {Kuhl}}]{hof24}%
	\BibitemOpen
	\bibfield  {author} {\bibinfo {author} {\bibfnamefont {T.}~\bibnamefont
			{Hofmann}}, \bibinfo {author} {\bibfnamefont {F.}~\bibnamefont {Schmidt}},
		\bibinfo {author} {\bibfnamefont {H.-J.}\ \bibnamefont {St\"ockmann}},\ and\
		\bibinfo {author} {\bibfnamefont {U.}~\bibnamefont {Kuhl}},\ }\href@noop {}
	{\bibinfo {title} {Spin resonance without a spin: A microwave analog}}
	(\bibinfo {year} {2024})\BibitemShut {NoStop}%
	\bibitem [{\citenamefont {Rehemanjiang}\ \emph {et~al.}(2020)\citenamefont
		{Rehemanjiang}, \citenamefont {Richter}, \citenamefont {Kuhl},\ and\
		\citenamefont {St{\"o}ckmann}}]{reh20}%
	\BibitemOpen
	\bibfield  {author} {\bibinfo {author} {\bibfnamefont {A.}~\bibnamefont
			{Rehemanjiang}}, \bibinfo {author} {\bibfnamefont {M.}~\bibnamefont
			{Richter}}, \bibinfo {author} {\bibfnamefont {U.}~\bibnamefont {Kuhl}},\ and\
		\bibinfo {author} {\bibfnamefont {H.-J.}\ \bibnamefont {St{\"o}ckmann}},\
	}\bibfield  {title} {\bibinfo {title} {Microwave realization of the chiral
			orthogonal, unitary, and symplectic ensembles},\ }\href
	{https://doi.org/10.1103/PhysRevLett.124.116801} {\bibfield  {journal}
		{\bibinfo  {journal} {Phys. Rev. Lett.}\ }\textbf {\bibinfo {volume} {124}},\
		\bibinfo {pages} {116801} (\bibinfo {year} {2020})},\ \Eprint
	{https://arxiv.org/abs/arXiv:1909.12886} {arXiv:1909.12886} \BibitemShut
	{NoStop}%
	\bibitem [{\citenamefont {Kottos}\ and\ \citenamefont
		{Smilansky}(1999)}]{kot99a}%
	\BibitemOpen
	\bibfield  {author} {\bibinfo {author} {\bibfnamefont {T.}~\bibnamefont
			{Kottos}}\ and\ \bibinfo {author} {\bibfnamefont {U.}~\bibnamefont
			{Smilansky}},\ }\bibfield  {title} {\bibinfo {title} {Periodic orbit theory
			and spectral statistics for quantum graphs},\ }\href
	{https://doi.org/10.1006/aphy.1999.5904} {\bibfield  {journal} {\bibinfo
			{journal} {Ann. Phys. (N.Y.)}\ }\textbf {\bibinfo {volume} {274}},\ \bibinfo
		{pages} {76} (\bibinfo {year} {1999})}\BibitemShut {NoStop}%
	\bibitem [{\citenamefont {Hofmann}\ \emph {et~al.}(2021)\citenamefont
		{Hofmann}, \citenamefont {Lu}, \citenamefont {Kuhl},\ and\ \citenamefont
		{St{\"o}ckmann}}]{hof21}%
	\BibitemOpen
	\bibfield  {author} {\bibinfo {author} {\bibfnamefont {T.}~\bibnamefont
			{Hofmann}}, \bibinfo {author} {\bibfnamefont {J.}~\bibnamefont {Lu}},
		\bibinfo {author} {\bibfnamefont {U.}~\bibnamefont {Kuhl}},\ and\ \bibinfo
		{author} {\bibfnamefont {H.-J.}\ \bibnamefont {St{\"o}ckmann}},\ }\bibfield
	{title} {\bibinfo {title} {Spectral duality in graphs and microwave
			networks},\ }\href {https://doi.org/10.1103/PhysRevE.104.045211} {\bibfield
		{journal} {\bibinfo  {journal} {Phys. Rev. E}\ }\textbf {\bibinfo {volume}
			{104}},\ \bibinfo {pages} {045211} (\bibinfo {year} {2021})},\ \Eprint
	{https://arxiv.org/abs/arXiv:2110.11722} {arXiv:2110.11722} \BibitemShut
	{NoStop}%
	\bibitem [{\citenamefont {Gnutzmann}\ \emph {et~al.}(2013)\citenamefont
		{Gnutzmann}, \citenamefont {Schanz},\ and\ \citenamefont
		{Smilansky}}]{gnu13}%
	\BibitemOpen
	\bibfield  {author} {\bibinfo {author} {\bibfnamefont {S.}~\bibnamefont
			{Gnutzmann}}, \bibinfo {author} {\bibfnamefont {H.}~\bibnamefont {Schanz}},\
		and\ \bibinfo {author} {\bibfnamefont {U.}~\bibnamefont {Smilansky}},\
	}\bibfield  {title} {\bibinfo {title} {Topological resonances in scattering
			on networks (graphs)},\ }\href
	{https://doi.org/10.1103/PhysRevLett.110.094101} {\bibfield  {journal}
		{\bibinfo  {journal} {Phys. Rev. Lett.}\ }\textbf {\bibinfo {volume} {110}},\
		\bibinfo {pages} {094101} (\bibinfo {year} {2013})}\BibitemShut {NoStop}%
	\bibitem [{\citenamefont {Guhr}\ \emph {et~al.}(1998)\citenamefont {Guhr},
		\citenamefont {M{\"u}ller-Groeling},\ and\ \citenamefont
		{Weidenm{\"u}ller}}]{guh98}%
	\BibitemOpen
	\bibfield  {author} {\bibinfo {author} {\bibfnamefont {T.}~\bibnamefont
			{Guhr}}, \bibinfo {author} {\bibfnamefont {A.}~\bibnamefont
			{M{\"u}ller-Groeling}},\ and\ \bibinfo {author} {\bibfnamefont {H.~A.}\
			\bibnamefont {Weidenm{\"u}ller}},\ }\bibfield  {title} {\bibinfo {title}
		{Random matrix theories in quantum physics: common concepts},\ }\href
	{https://doi.org/10.1016/S0370-1573(97)00088-4} {\bibfield  {journal}
		{\bibinfo  {journal} {Phys. Rep.}\ }\textbf {\bibinfo {volume} {299}},\
		\bibinfo {pages} {189} (\bibinfo {year} {1998})}\BibitemShut {NoStop}%
	\bibitem [{\citenamefont {Beenakker}(1997)}]{bee97}%
	\BibitemOpen
	\bibfield  {author} {\bibinfo {author} {\bibfnamefont {C.~W.~J.}\
			\bibnamefont {Beenakker}},\ }\bibfield  {title} {\bibinfo {title}
		{Random-matrix theory of quantum transport},\ }\href
	{https://doi.org/10.1103/RevModPhys.69.731} {\bibfield  {journal} {\bibinfo
			{journal} {Rev. Mod. Phys.}\ }\textbf {\bibinfo {volume} {69}},\ \bibinfo
		{pages} {731} (\bibinfo {year} {1997})}\BibitemShut {NoStop}%
	\bibitem [{\citenamefont {Bloembergen}\ \emph {et~al.}(1948)\citenamefont
		{Bloembergen}, \citenamefont {Purcell},\ and\ \citenamefont {Pound}}]{blo48}%
	\BibitemOpen
	\bibfield  {author} {\bibinfo {author} {\bibfnamefont {N.}~\bibnamefont
			{Bloembergen}}, \bibinfo {author} {\bibfnamefont {E.~M.}\ \bibnamefont
			{Purcell}},\ and\ \bibinfo {author} {\bibfnamefont {R.~V.}\ \bibnamefont
			{Pound}},\ }\bibfield  {title} {\bibinfo {title} {Relaxation effects in
			nuclear magnetic resonance absorption},\ }\href
	{https://doi.org/10.1103/PhysRev.73.679} {\bibfield  {journal} {\bibinfo
			{journal} {Phys. Rev.}\ }\textbf {\bibinfo {volume} {73}},\ \bibinfo {pages}
		{679} (\bibinfo {year} {1948})}\BibitemShut {NoStop}%
\end{thebibliography}
\end{document}